\documentclass[]{aa}  
\pdfoutput=1 
\usepackage{graphicx}
\usepackage{natbib}
\usepackage[varg]{txfonts}
\begin{document}
   \title{Measuring stellar differential rotation with high-precision space-borne photometry}
   \titlerunning{Stellar differential rotation}

   \subtitle{}

   \author{A.~F.~Lanza\inst{1} \and M. L. Das Chagas\inst{1,2} \and J. R. De Medeiros\inst{2}}
\authorrunning{A.~F.~Lanza et al.}

   \institute{INAF-Osservatorio Astrofisico di Catania, Via S.~Sofia, 78 -- 95123 Catania, Italy\\
              \email{nuccio.lanza@oact.inaf.it} \and
              Departamento de F\'{\i}sica, Universidade Federale do Rio Grande do Norte, Natal, RN, 59072-970 Brazil\\
              \email{renan@dfte.ufrn.br}
             }

   \date{Received ...; accepted ...}

\abstract{Stellar differential rotation is important for understanding hydromagnetic stellar dynamos, instabilities, and transport processes in stellar interiors, as well as for a better treatment of tides in close binary and star-planet systems.}{We introduce a method of measuring a lower limit to the amplitude of surface differential rotation from high-precision, evenly sampled photometric time series,  such as those obtained by space-borne telescopes. It is designed to be applied to main-sequence late-type stars whose optical flux modulation is dominated by starspots.}{An autocorrelation of the time series was used to select stars that allow an accurate determination of  starspot rotation periods. A simple two-spot model was applied together with a Bayesian information criterion to preliminarily select intervals of the time series showing evidence of differential rotation with starspots of almost constant area. Finally, the significance of the differential rotation detection and a measurement of its amplitude and uncertainty  were obtained by an a posteriori Bayesian analysis based on a  Monte Carlo Markov Chain  approach. We applied our method to the Sun and eight other stars for which previous spot modelling had been performed to compare our results with previous ones.}{We find that autocorrelation is a  simple method for selecting stars with a coherent rotational signal that is a prerequisite for successfully measuring differential rotation through spot modelling.  For a proper  Monte Carlo Markov Chain analysis, it is necessary to take the strong correlations among different parameters that exist in spot modelling into account. For the planet-hosting star Kepler-30, we derive a lower limit to  the relative amplitude of the differential rotation of $\Delta P / P = 0.0523 \pm 0.0016$. We confirm that the Sun as a star in the optical passband is not suitable for measuring differential rotation owing to the rapid evolution of its photospheric active regions. In general, our method performs well in comparison to more sophisticated and time-consuming approaches.}{}

   \maketitle
%

\section{Introduction}
\label{intro}

The Sun and other stars do not rotate as rigid bodies owing to latitudinal and radial transport of angular momentum induced by anisotropic turbulent Reynolds stresses, meridional flows, and magnetic fields \citep[e.g., ][]{Rudiger89}.
Differential rotation (hereinafter DR) plays a fundamental role in  hydromagnetic dynamo \citep{BrandenburgSubramanian05} and as a source of hydrodynamic and magnetohydrodynamic instabilities in stellar interiors \citep{KnoblochSpruit82}. 
Moreover, it plays a role in tidal interaction in close binary systems \citep{Scharlemann81,Scharlemann82} and is thus expected to affect the tidal interaction between a close-in planet and its host star \citep{Mathisetal13}. The measurement of the mean rotation period of a main-sequence late-type star, from which its age is estimated by the method of gyrochronology, is also affected by the amplitude of latitudinal DR \citep[see, e.g., ][]{EpsteinPinsonneault12}. 

In the Sun, we can study DR in detail in the photosphere by measuring the rotation rate at different latitudes by  Doppler shifts of the plasma spectral lines as well as by using sunspots as tracers for the motion of the surrounding plasma. The interior DR is accessible by helioseismic techniques that reveal a time dependence in some of the layers, probably related to the feedback of the Lorentz force associated with hydromagnetic dynamo action \citep[e.g., ][]{Howeetal00,Lanza07,Howe09}. 

In distant stars, we have much more limited information because only spatially unresolved data can be acquired. Recently, asteroseismic techniques have provided the first hints on radial DR in red giants \citep[e.g., ][]{Deheuvelsetal12,Mosseretal12,Goupiletal13}, while for main-sequence stars information on surface DR has been extracted through spectroscopic or photometric techniques. The advent of space-borne high-precision photometry with MOST \citep[Microvariability and Oscillations of STars experiment, ][]{Rucinskietal03}, CoRoT \citep[Convection, Rotation and Transit experiment, ][]{Auvergneetal09}, and Kepler \citep{Boruckietal10} has made available large and homogeneous datasets of photometric measurements of late-type stars that represent a treasure trove to study stellar rotation and DR, in particular. Therefore, we introduce in the present work a technique to measure stellar DR from high-precision and evenly-sampled photometric time series of late-type main-sequence stars and discuss its advantages and limitations in the context of previously proposed approaches. 

Main-sequence stars of the A and F spectral types are generally quite fast rotators and do not show brightness inhomogeneities in their photospheres 
{ \citep[see, however, ][]{Balona13}}, thus making  the effect of surface DR on the rotational broadening of spectral line profiles directly measurable by means of deconvolution techniques based on Fourier analysis \citep[e.g., ][]{ReinersSchmitt03}. Applying this approach, \citet{Reiners06} and \citet{AmmlervonEiffReiners12} measured DR in a sample of A and F stars and found a remarkable increase in its amplitude with increasing effective temperature \citep[see ][ for theoretical models that could explain such a dependence]{KukerRudiger05}. Stars of the spectral types G, K, and M generally show brightness inhomogeneities in their photospheres that are analogous to sunspots; i.e., they are associated with surface magnetic fields. Those having a sufficiently fast rotation ($ v \sin i \ga 15$~km~s$^{-1}$) can be mapped through the Doppler Imaging techniques \citep[e.g., ][]{Donatietal97,DonatiCollierCameron97,Strassmeier09}  allowing  their surface DR to be measured  and its dependence on temperature and rotation rate to be studied. \citet{Barnesetal05} find that the amplitude $\Delta \Omega$ of the DR has a weak dependence on the angular velocity of rotation $\Omega$, i.e., $\Delta \Omega \propto \Omega^{\alpha}$ with $\alpha = 0.15 \pm 0.10$, while a remarkably stronger correlation is found with stellar effective temperature, i.e.,  $\Delta \Omega$ increases with increasing effective temperature from M to G-type stars, thus extending the dependence found by \citet{Reiners06} to lower temperatures. { In main-sequence stars, the detected differential rotation is solar-like; i.e.,  the equator rotates faster than the pole. Nevertheless, an anti-solar differential rotation has been suggested in some late-type giants \citep[see, e.g., ][ and references therein]{Kovarietal07}}. 

For late-type stars that are slowly rotating ($v \sin i \la 12-15$~km~s$^{-1}$), Doppler Imaging cannot be applied and information on surface DR can be extracted solely by photometric techniques. The chromospheric fluxes in the cores of the Ca~II~H\&K lines have been monitored along several decades for a sample of late-type main-sequence stars in the framework of the Mt.~Wilson project to study rotation and stellar activity cycles. It has provided information on the dependence of  DR on stellar rotation rate \citep{Donahueetal96} thanks to the quite long lifetime of chromospheric plages in comparison with the stellar rotation period that makes them good tracers for pointing out the differences in rotation rate at different latitudes \citep{Donahueetal97a,Donahueetal97b}. 

In the Sun, the activity belts where active regions form migrate with the phase of the activity cycle. This is interpreted as a migration of the latitude of maximum toroidal magnetic field close to the base of the convection zone \citep[e.g., ][]{DikpatiCharbonneau99}. A similar migration is expected  in late-type stars that have a solar-like dynamo, producing a systematic variation in the period of the photometric modulation with the phase of the cycle. Such a variation has indeed been observed in the rotational modulation of the Sun-as-a-star chromospheric flux and provides an estimate of the amplitude of solar DR \citep{DonahueKeil95}. A key parameter  is the length of the time interval used to determine the seasonal solar rotation period. It is calibrated by trying to match  two contrasting requirements: a) avoid remarkable variations of the large scale pattern of chromospheric inhomogeneities that would imply using as short an interval as possible; b) attain sufficient time resolution and low false-alarm probability in determining the period of the rotational modulation that would benefit from a time interval that is as long as possible. In the Sun, the optimal extension of the seasonal time interval is found to be $150-200$ days that makes a compromise between the two opposite requirements. This is allowed because chromospheric active regions and activity complexes are remarkably long-lived in comparison with photospheric spots having a mean lifetime of $50-80$ days vs. $\sim 10-15$ days, respectively \citep[cf. ][]{Donahueetal97a,Lanzaetal03}. As a matter of fact, a similar approach based on photospheric sunspots was not successful because of the random longitude appearance and short lifetime of individual sunspot groups \citep{LaBonte82}. 

The situation is different in the case of very active, young solar-like stars whose rotation period is shorter than the Sun's and whose photospheric starspots have lifetimes of several months. Therefore, the method was successful  in that case \citep[see, e.g., ][]{MessinaGuinan02,MessinaGuinan03}. For the highly active and fast-rotating subgiant component stars in close binary systems such as RS Canum Venaticorum binaries, the persistence of active longitudes for decades allows us to measure a low-amplitude DR using photospheric starspots as tracers \citep[$\Delta \Omega /\Omega \sim 10^{-3}$ in HR 1099, cf. ][]{Lanzaetal06,BerdyuginaHenry07}. 

The majority of the stars observed by CoRoT and Kepler in the optical passband are not suitable to this approach because their photospheric active regions
have lifetimes that are shorter than the typical timescale of DR shear, i.e., $1/\Delta \Omega$, where $\Delta \Omega$ is the amplitude of the DR. This limits the precision in the determination of the rotation period attainable with periodogram techniques, even in the case of a uniformly sampled time series \citep{Lanzaetal93,Lanzaetal94}. On the other hand, if starspot intrinsic evolution is negligible, periodogram techniques coupled with a pre-whitening approach can be successful for estimating DR in solar-like stars by pinpointing the rotation frequencies of spots at different latitudes \citep{ReinholdReiners13,Reinholdetal13}. 

To make progress in the measurement of DR in late-type stars having starspots that evolve on a timescale comparable to $1/\Delta \Omega$ or possibly shorter, we investigate the potentiality of a simple starspot model to extract DR. Our approach applies a simple autocorrelation technique to estimate the coherence time of the light modulation that provides an estimate of the spot lifetime to be compared with the shear timescale. This allows us to select  promising candidates for spot modelling. For a given star, we perform a screening of the time intervals showing variations that  likely stem from the effect of DR rather than from intrinsic starspot  evolution. Finally, we apply a Monte Carlo Markov Chain (hereafter MCMC) method to estimate the most probable value of DR and its standard deviation following an  approach introduced by \citet{Croll06}. 
We compare the proposed method with previous ones by analysing a sample of stars observed by MOST, CoRoT, and Kepler whose DR has been extracted with different spot modelling approaches. Moveover, we also consider the case of the Sun as a star to show the limitation of the method in the case of a slowly rotating star.

\section{Observations}
\label{observations}

\begin{figure*}
\centering{
\includegraphics[width=14cm,height=24cm,angle=0]{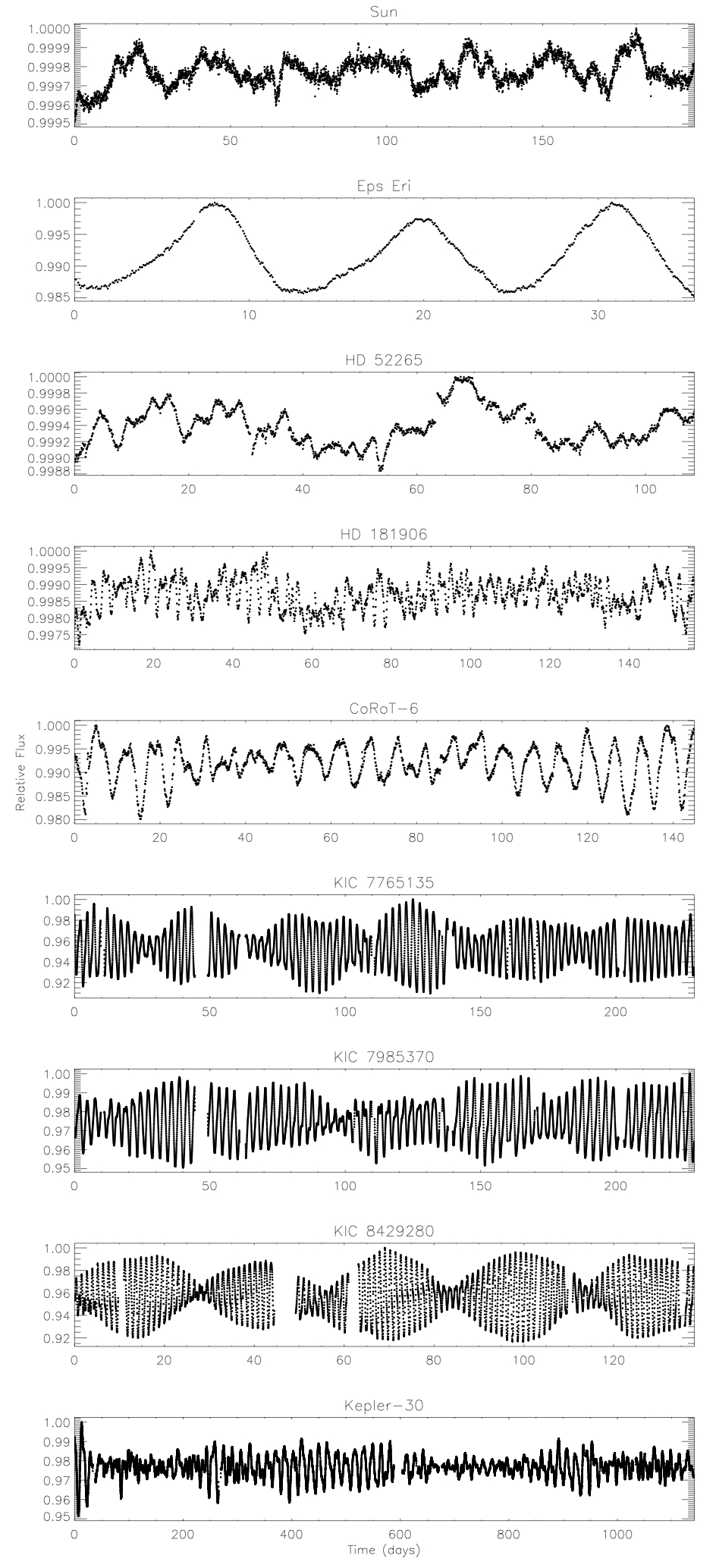}} 
\caption{Photometric time series of the Sun and the other considered stars. The flux has been normalized to the maximum value observed along each  time series. }
\label{lc_plots}
\end{figure*}

Aiming at a comparison of our new approach with previous estimates of DR, we consider the Sun and eight distant stars for which high-precision space-borne photometry is available. In the case of the Sun, we use a total solar irradiance (hereafter TSI) time series acquired by the VIRGO experiment on board the SoHO satellite \citep{Froehlichetal95,Froehlichetal97a,Froehlichetal97b}. 
We use Level 2 data consisting of measurements  acquired in a bolometric passband (irradiance) with a cadence of one hour, reduced to a fixed distance of 1 AU and corrected for the degradation of the instrument exposed to the space environment and other short- and long-term systematics by comparison with other radiometers as explained on the experiment's web page \footnote{http://www.pmodwrc.ch/pmod.php?topic=tsi/virgo/proj\_space\_virgo}, from which the data have been downloaded. The variation in the TSI is dominated by photospheric active regions with sunspot groups producing flux dips during their transit across the solar disc, while faculae produce an increase in the flux when they are close to the limb since their contrast is at its maximum there, while it is negligible close to disc centre. The diffused magnetic network provides an additional contribution that is only modestly modulated with solar rotation \citep[e.g., ][]{Fliggeetal00}. 
The lifetime of sunspot groups is generally shorter than one rotation, and the TSI modulation they produce is often dominated by their intrinsic evolution making it difficult to use them to measure solar rotation period \citep[see, e.g., ][]{Lanzaetal03,Lanzaetal04}. 
The modulation produced by faculae is generally more coherent up to 70-100 days, i.e., for three to four rotations. They dominate the TSI variability close to the minimum of the eleven-year cycle. Therefore, we select a time interval of 200 days starting from 30 November 1996 during which the rotational modulation signal is most evident and sizable sunspot groups are not detected, that is no clear light dip is observed in the TSI.  This time interval provides us with the best rotational signal of the Sun as a star (see Fig. \ref{lc_plots}, upper panel). The relative precision of  individual measurements of the TSI is $\sim 20$ parts per million (hereafter ppm), i.e., comparable to the precision of Kepler stellar photometry for a G-type star of apparent visual magnitude  $\sim 12$ binned at an exposure time of $\sim 1$ hour.

In addition to the Sun, we considered eight other stars.  
For the K2 main-sequence star \object{$\epsilon$ Eridani}, a time series of 35.495 days was acquired by MOST  \citep{Walkeretal03} 
starting on 28 October 2005.  The data were binned at the orbital period of the satellite of 101.41 minutes after removing measurements affected by  bad pointing, high background/stray light, and  proton hits on the CCD during the crossing of the South Atlantic Anomaly of the Earth's magnetic field \citep{Crolletal06}. 
The intrinsic point-to-point precision of the light curve is $\sim 50$ ppm. Reduced data have been downloaded from the mission public data archive\footnote{http://most.astro.ubc.ca//data/data.html}. Before our analysis, a long-term linear trend was subtracted to remove residual instrumental or uncorrected background effects.

Among the targets observed by CoRoT \citep{Auvergneetal09}, 
we considered the two asteroseismic targets \object{HD~52265} \citep{Ballotetal11} and \object{HD~181906} \citep{Mosseretal09} for which previous estimates of DR were obtained with starspot models and some information on the inclination of the rotation axis has been provided by modelling the rotational splitting of the p-mode oscillation frequencies. 

HD~52265 is a G0 main-sequence star hosting a planet with an orbital period of 119~days and a projected mass of 1.13 Jupiter masses \citep{Butleretal00}. It has been observed by CoRoT for 117 days starting from 13 November 2008. We  downloaded the reduced  level-2 data  in the heliocentric time frame (the so-called Helreg level 2 data) with an original cadence of 32~s from the CoRoT Public Data Archive\footnote{http://idoc-corot.ias.u-psud.fr/}. The data were binned at the orbital period of the satellite of 6184~s, and outliers and a long-term trend were removed as described in Sect.~2.1 of \citet{DeMedeirosetal13}.  The same procedure was applied to the time series of HD~181906, an F8 main-sequence star for which a  data set of 156.6~days was  acquired by CoRoT starting from 11 May 2007. For these two targets observed in the asteroseismic field of CoRoT focal plane, a relative precision of $\approx 10$~ppm for  individual binned measurements is achieved.  

We also apply our approach to the light curve of \object{CoRoT-6}, an F9 main-sequence star accompanied by a transing planet with an orbital period of 8.886~days and a mass of $\sim 2.9$ Jupiter masses. It is remarkably active, and its photospheric spots have been mapped by \citet{Lanzaetal11} deriving an estimate of its mean rotation period and  amplitude of  DR. This star was  observed in the exoplanet field of the CoRoT focal plane for 144.9 days starting from 15 April 2008 with an original cadence of 32~s. The level 2 photometry was binned at the orbital period of the satellite, while transits, outliers, and long-term trends were removed as described in Sect.~2 of \citet{Lanzaetal11}. Since CoRoT-6 is much fainter than CoRoT asteroseismic targets, the relative standard error of the binned points is  $\sim 240$~ppm.    

Finally, we consider four Kepler targets whose DR and spot activity have previously been studied.
\object{KIC~7765135} and \object{KIC~7985370} are young Sun-like stars (spectral type G1.5V) investigated by \citet{Froehlichetal12}. We have considered Kepler photometry since 2~May~2009 extending for 228.9~days with a cadence of  29.4~minutes. The relative precision of each point as derived from the photon shot noise is   $\sim 40$ and $\sim 18$~ppm, respectively. \object{KIC~8429280} is a young main-sequence star (spectral type K2) studied  by \citet{Frascaetal11}, for which we considered a data set of 137.9 days starting from 2 May 2009 with the same 29.4 min cadence and a relative accuracy of the individual data points of $\sim 17$~ppm. 

\object{Kepler-30} is a Sun-like star accompanied by three transiting planets \citep{Fabryckyetal12}. Starspot occultations detected during planetary transits have allowed the projected obliquities of the stellar spin axis  to the orbital planes of its planets to be constrained \citep{SanchisOjedaetal12}. In turn, this allowed us to constrain the inclination of the stellar spin axis to the line of sight  to improve the spot modelling of the out-of-transit light curve. The dataset we consider consists of 1141.5 days of observations starting from 13 May 2009 with a cadence of 29.4 min. The relative accuracy of each data point is  $\sim 260$~ppm. { For our analysis, planetary transits were removed according to the ephemeris given in Table~1 of \citet{SanchisOjedaetal12}. We also visually checked that the obtained out-of-transit light curve  has no evident transit signal left over. The introduced gaps do not exceed  $0.35$~days (i.e., the duration of the transit of the most distant planet Kepler-30d) and  have a negligible effect on our spot modelling because the rotation period of the star is $\sim 16$~days and the mean evolutionary timescale of the spot pattern is $\sim 22$~days (cf. Sects.~\ref{autocorr_analysis} and~\ref{subinterval-selection}). }

All the light curves of these four stars were downloaded from the Kepler data archive\footnote{http://keplergo.arc.nasa.gov/DataAnalysisRetrieval.shtml}. 
To account for long-term trends of instrumental origin and the effects of the re-orientation of the spacecraft after each $\sim 90$ days (a time interval called "a quater" in the Kepler jargon), Kepler archive provides  the so-called co-trending basis vectors \citep[e.g., ][]{Twickenetal10}. They specify  trends that are common to stars close to each other on the Kepler focal plane, thus accounting for most of the common instrumental effects. However, instrumental effects specific to a given target, e.g., depending on background contamination at its specific location or on other subtle effects \citep[see, e.g., ][]{Basrietal11}, cannot be accounted for. 

A linear combination of co-trending basis vectors can be subtracted from the raw time series to obtain de-trended data. Nevertheless, we have not applied this   approach because our analysis will focus on time intervals comparable to one stellar rotation, i.e., ranging from a few to $\sim 15-20$ days at most. To detect discontinuities and outliers, and de-trend the data on such short timescales, the method developed by \citet{DeMedeirosetal13} has a comparable performance. It is based on a simple algorithm to identify discontinuities and correct for the long-term trend by fitting a third-order polynomial, similar to the approach of \citet{Basrietal11}. In consideration of its  generality and simplicity,  we decided to apply it  to our Kepler time series. The complete set of analysed photometric time series is plotted in Fig.~\ref{lc_plots}.

\section{Light curve modelling and analysis}

\label{overview}

Our  approach to detecting and estimating surface DR has been designed by considering  application to fairly large samples of stellar time series. As a first step, we look for those stars having the most stable signal in terms of rotational modulation because they are the best candidates for estimating the mean rotation period, as well as DR. 
We quantify the stability of the signal by means of  the autocorrelation of its time series. For  targets with a sufficiently stable signal, we apply spot modelling to derive individual spot rotation periods. To keep the model as simple as possible, we only consider two spots to model the flux modulation. Since starspots evolve, we cut the time series into equal intervals of length $\Delta T$ during which spots can be assumed to remain stable. To find $\Delta T$, we consider shorter and shorter intervals until the  best fit obtained with non-evolving spots  becomes acceptable (see below). For each of these intervals, we compare spot models with and without DR, i.e., we let the two spots  have different or the same rotation period, respectively, and compare the goodnesses of  fit obtained in the two hypotheses. In this way, we find the time interval for which the assumption of two different rotation periods, i.e., DR, gives the best improvement over the hypothesis of rigid rotation. For that interval, the a posteriori distributions of the rotation periods of  the spots are determined using an MCMC approach as in \citet{Croll06}, to have a sound statistical estimate of the amplitude of DR and its uncertainty. 

The philosophy behind our approach is that not all the intervals of a long time series are equally suited to providing a  measure of DR because its signal can be hidden by the intrinsic evolution of starspots. Therefore, we first look for stars that have a  stable rotational modulation signal, as quantified in Sect.~\ref{autocorrelation} below. Then we seek  the interval(s) with the best fit in terms of spots with a fixed area and DR, i.e., during which the impact of starspot evolution is the weakest  and a significant DR signal appears to be present. Even if this interval covers only one or two rotations of the star, previous studies by, say, \citet{Crolletal06}, \citet{Croll06}, and \citet{Froehlich07} have demonstrated that spot modelling can extract a meaningful signal of DR. This is possible thanks to the sensitivity of spot modelling to the drift in longitude of  individual spots produced by a latitudinal shear. Even  drifts as small as $20^{\circ}-30^{\circ}$ per rotation period  can be significantly detected  \citep[e.g., ][]{Lanzaetal07,SilvaValioLanza11}. On the other hand, techniques based on periodogram analysis need larger phase shifts, up to $180^{\circ}$,  to resolve the peaks corresponding to the rotation frequencies of individual spots \citep{Lanzaetal03,Lanzaetal04,ReinholdReiners13}. Therefore, they are  prone to severe problems owing to the intrinsic spot evolution because such large shifts are produced only after a timescale comparable to $1/\Delta \Omega$, where $\Delta \Omega$ is the amplitude of DR. 
In the following sections, we describe the successive steps in our approach in detail.

\subsection{Autocorrelation of photometric time series}
\label{autocorrelation}

The coherence of the rotational modulation signal can be quantified by the relative heights of successive peaks in the autocorrelation function of the time series. The autocorrelation function provides a good estimate of the mean stellar rotation period in the case of evenly sampled time series extending for several rotation periods, as shown by \citet{McQuillanetal13}, among others. Assuming that the  rotational modulation of the flux is produced by only two spots that do not evolve in time, the flux can be expressed as 
\begin{equation}
F(t)  = \sum_{j=1}^{2} \sum_{k=0}^{\infty} \alpha_{kj} \cos (k \Omega_{j} t +\phi_{k j}),
\end{equation}
where we have developed the contribution of each spot in a Fourier series since it is a strictly periodic function of the time $t$; $\Omega_{j} \equiv 2\pi/P_{j}$ is the angular velocity of  the $j$-th spot with $P_{j}$ being its rotation period, $\alpha_{kj}$ the Fourier coefficient of order $k$ for spot $j$, and $\phi_{kj}$ its initial phase.  The flux modulation vs. time has a continuous first derivative \citep[cf. ][]{Russell1906}, therefore the Fourier coefficients of the above series decrease with increasing order as $k^{-2}$ \citep{Smirnov64}, implying that only the first terms, say, up to $k=2-4$, are relevant for describing the modulation \citep[cf. ][]{Cowanetal13}. 

In the case of a continuous, indefinitely extended signal having zero mean, the normalized autocorrelation function can be defined as 
\begin{equation}
A(\tau) \equiv \lim_{T \rightarrow \infty } \frac{L(\tau)}{L(0)}, 
\end{equation}
where
\begin{equation}
L(\tau) \equiv \int_{-T/2}^{T/2} F(t) F(t+\tau) dt,
\label{eqL}
\end{equation}
and $\tau$ is the time lag. In practice, our time series has a finite extension $T_{\rm max}$, so we  consider a finite interval of integration assuming that 
$|\tau| < T_{\rm max}/2$. If our two spots have the same rotation period $P= P_{1} = P_{2}$, the autocorrelation oscillates with a period $P$ showing  equal maxima at $\tau = n P$, where $n$ is an integer. When the intrinsic evolution of starspots is negligible, all Fourier coefficients $\alpha_{kj}$ are constant, and we can compute the autocorrelation for $P_{1} \not= P_{2}$ considering an integration extended to a time interval $[-T/2, T/2]$, where $T \gg \Omega_{1}^{-1}, \Omega_{2}^{-1}, | \Omega_{1} -\Omega_{2}|^{-1}$. After some manipulations, we obtain 
\begin{equation}
A(\tau) = \frac{\sum_{k} \left\{ 2 \alpha_{k1}^{2} \cos \left[ k \left( \frac{\Omega_{1}-\Omega_{2}}{2} \right)\tau \right] +\left( \alpha_{k2}^{2} - \alpha_{k1}^{2}\right)  \right\} \cos \left( k \hat{\Omega} \tau \right)}{ \sum_{k} \left( \alpha_{k1}^{2} + \alpha_{k2}^{2} \right)},
\end{equation}
where $\hat{\Omega} = (\Omega_{1} + \Omega_{2})/2$ is the mean angular velocity. In other words, $A(\tau)$ is a periodic function of period $2 \pi/\hat{\Omega}$ whose amplitude is modulated  with a beating period $ 4 \pi/ |\Omega_{1} -\Omega_{2}| $. The separation between successive minima of the amplitude is half that period. When the rotation is rigid, i.e., $\Omega_{1} = \Omega_{2} = \hat{\Omega}$, we obtain: $A(\tau) \propto \cos (k \hat{\Omega} \tau )$. 

We can predict the effects of  starspot evolution in a simple case, i.e., by assuming that the time dependence of  Fourier coefficients obeys the relationship $\alpha_{kj} (t +\tau) = f(\tau) \alpha_{kj} (t)$ for all values of $t$ and $\tau$. Such a specific time dependence, associated with the initial condition $f(0)=1$, gives $f(\tau) = \exp(-\tau / \tau_{\rm s})$, where $\tau_{\rm s}$ is the starspot lifetime that we assume to be a constant. In other words, in the above hypothesis, we have an exponential decay of starspot area. This is approximately valid for non-recurrent sunspots \citep{Bumba63}, although their decay rate  varies widely from one active region to the other \citep[for a recent review see, e.g.,  ][]{MartinezPilletetal93}. Considering an integration interval with $T \gg \Omega_{1}^{-1}, \Omega_{2}^{-1}, | \Omega_{1} -\Omega_{2}|^{-1}$, we obtain  
\begin{eqnarray}
\lefteqn{A(\tau)   =  \exp (-\tau / \tau_{\rm s}) } \\ \nonumber 
& &  \times  \; \frac{\sum_{k} \left\{ 2 \alpha_{k1}^{2}(\bar{t}) \cos \left[ k \left( \frac{\Omega_{1}-\Omega_{2}}{2} \right)\tau \right] +\left[ \alpha_{k2}^{2}(\bar{t}) - \alpha_{k1}^{2}(\bar{t})\right]  \right\} \cos \left( k \hat{\Omega} \tau \right)}{ \sum_{k} \left[ \alpha_{k1}^{2}(\bar{t}) + \alpha_{k2}^{2}(\bar{t}) \right]},  
\label{auto_general}
\end{eqnarray}
where $\bar{t}$ is a time corresponding to the mean value of the Fourier coefficients along the  interval $[-T, T]$. When the starspot lifetime $\tau_{\rm s}$ is shorter than the beating period $4 \pi / |\Omega_{1} - \Omega_{2}|$, Eq.~(\ref{auto_general}) indicates that the autocorrelation function $A(\tau)$ has a sequence of relative maxima whose amplitude decreases exponentially with time lag as $\exp (-n \hat{P}/ \tau_{\rm s})$, where $\hat{P} \equiv 2 \pi/\hat{\Omega}$ is the mean rotation period of the spots. This allows us to estimate the characteristic decay time of starspots that is  fundamental  to establishing whether they  are suitable tracers to measure DR. On the other hand, when $\tau_{\rm s} \gg 4 \pi / |\Omega_{1} - \Omega_{2}|$,  intrinsic starspot evolution can be neglected, and the beating period of the autocorrelation function can be used to estimate the amplitude of  DR. 
{ Of course, other methods can be applied to estimate the typical lifetime of starspots from the changing modulation of a photometric time series, thus providing further information on $\tau_{\rm s}$ \citep[cf. ][]{Lehtinenetal11,Lehtinenetal12}.}
 
Real time series are affected by noise that  produces spurious peaks in the autocorrelation. For a normally distributed uncorrelated random variable, the expectation value of $A(\tau) \simeq 1/N$, where $N$ is the number of data points in the series and its variance is $\sigma^{2} = 1/N$. Therefore, a peak can be considered to be significant at a 2$\sigma$ level when its amplitude exceeds $\sim 2/\sqrt{N}$. Usually, a correction is applied to this formula to account for a possible short-range correlation among the values of the random variable as expected when a physical process with some degree of self-correlation is responsible for the fluctuations. We adopt the so-called large-lag approximation to estimate the variance of $A(\tau)$ \citep{Anderson76}. In the case where $\tau = m \Delta t$, i.e.,  considering the autocorrelation at evenly spaced values of the lag with successive values separated by $\Delta t$, 
\begin{equation}
\sigma^{2}(\tau) = \sigma^{2} (m \Delta t) = \left( 1 + 2 \sum_{p=0}^{m-1} [A (p \Delta t)]^{2} \right) \frac{1}{N}; 
\end{equation}
that is, the variance of the autocorrelation at lag $m \Delta t$ takes the autocorrelation at all the lags shorter than $m \Delta t$ into account. This formula makes the variance of the autocorrelation  increase with increasing lag in comparison to the case of a completely uncorrelated random variable. 

We use the IDL function {\tt A\_CORRELATE} to compute the autocorrelation function. It does not consider any gaps along the time series. Nevertheless, their impact is very small because of the almost perfect duty cycle of space-borne observations ($ \ga 95$ percent), therefore, we shall not apply algorithms developed for time series with uneven sampling \citep[e.g., ][]{EdelsonKrolik88}. 

\subsection{Spot modelling}
\label{spot_modelling}

We have adopted a simple spot model to fit the light modulation of late-type stars in order to keep the number of free parameters as small as possible. This is advantageous when applying the MCMC method (see below). The flux of the star is written as
\begin{equation}
F(t) = F_{0} + \sum_{j=1}^{2} \Delta F_{j}(t),
\end{equation}
where $F_{0}$ is a  constant value and $\Delta F_{j}$  the flux variation due to the $j$-th spot. We assume that $F_{0}$ may be different from the flux of the unspotted star to account for the  effect of several small active regions evenly distributed in longitude, in addition to the two discrete spots responsible for the flux modulation.  The specific intensity over the disc of the unperturbed star is specified by a quadratic limb-darkening law as 
\begin{equation}
I(\mu) = a_{\rm p} + b_{\rm p} \mu + c_{\rm p} \mu^{2},
\end{equation}
where $a_{\rm p}$, $b_{\rm p}$, and $c_{\rm p}$ are  limb-darkening parameters that depend on the effective temperature, gravity, and chemical abundance of the stellar atmosphere, while $\mu \equiv \cos \psi$, where $\psi$ is the angle between the normal to the given surface element and the line of sight. The unperturbed flux coming from the stellar disc is $F_{\rm U} = \pi R_{*}^{2} (a_{\rm p} + 2b_{\rm p}/3 + c_{\rm p}/2)$, where $R_{*}$ is the radius of the star.

For simplicity, we consider only dark spots, although  photospheric faculae may be relevant for stars with a level of activity comparable to the Sun's  \citep{Lanzaetal03,Lanzaetal04}, while for more active stars, they have a minor impact on the rotational modulation of the flux \citep[see ][ and references therein]{Gondoin08,Lanzaetal09,Lanzaetal09a}. The specific intensity in a spot is assumed to be $I_{\rm s} (\mu) = (1- c_{\rm s}) I(\mu)$, where the 
spot contrast $c_{\rm s}$ is assumed to be constant for a given star. In this way, the relative flux variation due to the $j$-th spot is
\begin{equation}
\frac{\Delta F_{j}}{F_{\rm U}} = - c_{\rm s }\left( \frac{a_{j}}{\pi R_{*}^{2}} \right) \left(\frac{a_{\rm p} + b_{\rm p} \mu + c_{\rm p} \mu^{2}}{a_{\rm p} +\frac{2}{3} b_{\rm p} +\frac{1}{2} c_{\rm p}} \right) v(\mu)\, \mu,  
\label{spot_effect}
\end{equation}
where the value of $\mu$ at time $t$ for the $j$-th spot is given by
\begin{equation}
\mu = \sin i \sin \theta_{j} \cos[\lambda_{j} + \Omega_{j}(t-t_{0})] + \cos i \cos \theta_{j},
\label{mu_def}
\end{equation}
where $a_{j}$ is the area of the $j$-th spot, $\lambda_{j}$ and $\theta_{j}$  its longitude and colatitude, $i$  the inclination of the stellar spin axis to the line of sight, and $\Omega_{j} = 2 \pi/P_{j}$ the angular velocity of rotation of the spot, with $t_{0}$ the initial time. The visibility of the spot $v(\mu)$ is equal to 1 when $\mu > 0$ and is zero otherwise; for simplicity, we can express it as $v(\mu) = (\mu + |\mu |)/2$. This model is valid when the area of the spot is much smaller than the area of the stellar disc; i.e., we assume point-like spots that is justified for stars having an activity level not remarkably greater than that of the Sun where the largest sunspot groups reach a few $10^{-3}$ of the hemisphere. 

The above model is applied to individual intervals of the time series of length $\Delta T$ to reduce the impact of the evolution of the spots. Specifically, we consider a subdivision of a time series of length $T$ into $M$ intervals of equal length $\Delta T  = T /M$. The optimal value is found by trial and error by increasing $M$ until we obtain an acceptable fit. In the case of the Sun, the optimal length is  $\sim 14$ days \citep{Lanzaetal03,Lanzaetal07}, but it varies widely from one star to the other. 

For a given interval, our model has ten free parameters consisting of the unmodulated flux $F_{0}$, the inclination of the stellar spin axis $i$, the area and coordinates of the two spots ($a_{j}, \theta_{j}, \lambda_{j}$, with $j=1,2$), and their rotation periods $P_{1}$ and $P_{2}$. During the search for the best value of $M$, we usually fix the inclination to avoid strong correlations between it and the spot areas and colatitudes that can lead to bad fits. The observed light curve is normalized to the maximum value of the flux observed along the whole time interval $T$ before fitting the relative flux variations with our model. A Levenberg-Marquardt algorithm is applied to minimize the chi square after fixing the allowed ranges of variation of the parameters to avoid unphysical results, e.g., negative spot areas. This is possible using, say, the IDL routine MPFIT \citep{Markwardt09}\footnote{See also http://purl.com/net/mpfit}. Since the algorithm explores the chi square landscape starting from an initial point, the choice of that point is critical for converging to a good solution. In other words, if the initial point is too  far from the one giving the best fit, the algorithm can get stuck at a local minimum providing a poor fit. 
Therefore, we estimate initial values of the spot areas and longitudes for a given inclination from the depths and times of the light minima along the given  interval. Then we compare the minimum chi squares obtained with different initial longitudes of the spots by varying them by multiples of $90^{\circ}$ with respect to the initial longitudes estimated from the times of light minima. In this way, we select the initial values of the parameters leading to the best fit. 

To model the light curve with rigid rotation, we set $P_{1}=P_{2}$ in our model and compute the best fit using the one corresponding to the previous best fit with $P_{1} \not= P_{2}$, i.e., allowing for DR, as a starting point. This generally ensures convergence to an acceptable fit.

\subsection{Looking for the intervals with the best DR signal}
\label{interval_selection}

For each interval of the light curve of duration $\Delta T$, we compare the chi squares obtained with and without DR, usually adopting a fixed inclination of the stellar spin axis to reduce  degeneracies among parameters that often produce convergence problems. Since two spots are generally not enough to fit a light curve down to the photometric precision \citep[cf.  ][]{Lanzaetal03,Lanzaetal07}, a component of the residuals is certainly associated with small spots not considered in our model. We assume that they evolve on a timescale significantly shorter than $\Delta T$, so we can treat their effect as an increase in the random noise as a crude approximation. The chi square  of the best fit obtained with DR, $\chi^{2}_{\rm DR}$, is generally smaller than that obtained with the rigidly rotating model, $\chi^{2}_{\rm RR}$ that has only nine free parameters instead of ten. Therefore, we adopt the Bayesian information criterion (hereafter BIC) to measure the relative goodness of fit of one model vs. the other \citep[e.g., ][]{Liddle07}. The difference in the BIC estimator between the two models is
\begin{equation}
\Delta {\rm BIC} = \Delta \ln {\cal L} - \ln N,
\label{delta_bic}
\end{equation}
where $\Delta {\cal L}$ is the difference in the likelihood function of the two models and $N$ the number of data points in the fitted time interval. The second term in Eq. (\ref{delta_bic}) accounts for the difference in the number of free parameters in the two models. Since a two-spot model is generally not able to fit a light curve down to the precision of the measurements, we consider the standard deviation of the residuals of the best fit obtained with the DR model as the "true" standard deviation of the random noise present in the data, according to the above hypothesis on the photometric effects of small spots not accounted for in our model. This is equivalent to scaling the standard deviation in the definition of  $\chi^{2}_{\rm DR}$ so as to obtain  $\chi^{2}_{\rm DR}=N$.

Assuming that the deviations from the model are random and normally distributed, the likelihood is proportional to $\exp (-\chi^{2})$ \citep[e.g., ][ Ch. 15.2]{Pressetal02}.  Therefore, the difference in the logarithm of the likelihood function is
\begin{equation}
\Delta \ln {\cal L} = 2 \left( \frac{\chi^{2}_{\rm RR}}{\chi^{2}_{\rm DR}} -1 \right). 
\label{delta_likely}
\end{equation}
We regard  a value $\Delta {\rm BIC} \geq 2$ as preliminary evidence in favour of the model with DR. As a matter of fact,  BIC is an asymptotic approximation to a full Bayesian model comparison. It is rigorously valid when the noise is uncorrelated and identically distributed in a normal (or at least exponential) way all along the fitted interval. This is not rigorously valid in our case because the photometric effects of small spots are correlated on their evolution timescale,  and they may not be uniformly distributed along the analysed data set. Therefore, we use the difference in the BIC estimator as only a preliminary indication for selecting those subintervals having the best evidence  of a DR signature because a full Bayesian analysis of the entire time series would be computationally prohibitive even for a single star (see below). 

Our procedure can be summarized as follows. Firstly, the best fits for intervals $\Delta T$ obtained for  different values of $M$ and with and without DR are visually inspected when $\Delta {\rm BIC} \geq 2$ to remove cases in which the difference in the $\chi^{2}$ values is due to convergence problems (usually happening in 15-20 percent of the cases). Secondly, among the intervals with proper best fits, we choose the one having the clearest signal as revealed by the improvement in the reproduction of the times of light minima and/or the variation in the amplitude of the modulation when DR is included. Finally, this interval is  considered for a detailed MCMC analysis. 

\subsection{Extracting DR with an MCMC approach}
\label{mcmc_method}

We apply the MCMC approach proposed by \citet{Croll06} to estimate the mean values of the rotation periods $P_{1}$ and $P_{2}$ of the two individual spots, and derive the a posteriori distributions of $P_{1}$ and $P_{2}$ to be used to assess the significance of the DR.  We use the Metropolis-Hasting algorithm to generate a chain of random points in the parameter space to sample the a posteriori distribution of the parameters \citep[see, e. g.,][ Ch. 15.8]{Pressetal02}. Since  correlations among different parameters are very strong (see Sect.~\ref{param_corr}), the ergodicity of the obtained chain is generally not guaranteed and must be checked a posteriori. One way to reduce correlations among the parameters is to accept only those points that correspond  to values of the $\chi^{2}$ close to the minimum $\chi^{2}$. We generally adopt a threshold between one and five percent above the minimum $\chi^{2}$ for acceptance. Apart from such a constraint, the Metropolis-Hasting algorithm is  applied in its standard form. We indicate a generic point in the parameter space as  ${\bf Q} = \{ q_{1}, q_{2}, ... q_{10} \} $, where $q_{s}$ are the individual free parameters of the model, with $s=1,2, ...,10$, because we generally also allow the inclination to vary. The starting point of our chain in the parameter space is the one corresponding to the minimum $ \chi^{2} $ in the case of the two-spot model with DR. Sometimes, we found that running the Metropolis-Hasting algorithm for $ \sim 10^{5}-10^{6}$ steps a significantly  lower $\chi^{2}$ value is found. In this case, we adopted the new minimum as the starting point. To generate the next step of the chain  from the current point ${\bf Q}_{k}$, we generate a normal random deviate for each parameter $q_{s}$  to obtain a candidate point ${\bf Q}_{k+1}$. If the corresponding value of the $\chi^{2}$ is lower than that of the current point, the new point is accepted as the next step in the chain, otherwise we accept it with  probability  $\exp [-(\chi^{2}_{k+1} -\chi^{2}_{k})/\chi^{2}_{\rm min}]$, where $\chi^{2}_{k}$ is the chi square of the fit obtained with the parameter set ${\bf Q}_{k}$ and $\chi^{2}_{\rm min}$  the chi square corresponding to the minimum from which the chain is started. We adjust the standard deviations of the normal deviates of the individual parameters in order to have an overall acceptance rate between 20 and 30 percent, as recommended by \citet{Croll06}. Since we compute very long chains, i.e., with at least $40-60$ million points, we can easily obtain shorter chains for checking mixing and convergence of the MCMC procedure simply by cutting a long chain into subchains. Following \citet{Croll06}, we usually consider four subchains after discarding the initial $\sim 10^{6}$ steps that represent the so-called burning phase of the chain. We also apply a thinning factor of 10; i.e., we take only one point out of ten consecutive points to remove  local correlations in the subchains. 
The mixing and convergence of the subchains, i.e., their ergodicity,  is tested by applying the method of Gelman and Rubin as implemented in Sect.~3.2 of \citet{Verdeetal03}. We consider a subchain as successfully converged and mixed when the parameter $R$ of Gelman and Rubin is lower than 1.2 for all the model parameters \citep[see ][for details]{Verdeetal03,Croll06}. 

\subsection{Parameter correlations in spot modelling}
\label{param_corr}

In several cases, we found that it is not possible to reduce $R$ below the acceptance threshold of 1.2 for all the parameters even after running chains with $ 5 \times 10^{8}$ points. Sometimes, this is due to the miss of the best minimum of the chi square by the Levenberg-Marquardt algorithm. In some cases, by running the Metropolis-Hasting algorithm for several million points, we are able to improve the minimum and the convergence of the chain as found in the case of CoRoT-6. However, in several other cases, this is not sufficient.  
An inspection of the  parameter values along the  chain reveals that this occurs when there are strong correlations  among the parameters. Since a light curve is a one-dimensional data set, it only contains very limited information to constrain a two-dimensional spot map \citep[cf. ][]{Cowanetal13}. If the unspotted reference level is fixed as in our model, the epochs of the light curve minima provide information on spot longitude, while the depths of the minima give information on the projected spot area. Therefore, the inclination of the stellar spin axis $i$, the colatitude of a given spot $\theta_{j}$, and its area $a_{j}$ are strongly degenerate, as one can see from Eqs.~(\ref{spot_effect}) and (\ref{mu_def}).  When the convergence of a chain is not achieved as a consequence of these degeneracies, we can try different strategies. The first is to introduce constraints that take  the correlations between the inclination $i$ and the colatitudes of the spots $\theta_{j}$ into account as well as between the initial longitudes $\lambda_{j}$ and the angular velocities $\Omega_{j} \equiv 2 \pi/P_{j}$. Specifically, we impose $\delta \theta_{j} = \delta i$ and $\delta \lambda_{j} / \lambda_{j} = - \delta P_{j}/P_{j}$, where $\delta$ indicates the variation in the given quantity in a candidate step of the chain and $j=1,2$ indicates the spot (see Appendix~\ref{app_corr} for a justification of these correlations). When such constraints are not enough, we may calculate a linear regression a posteriori between the parameter with  the largest $R$ and the other free parameters to account for further correlations. Specifically, in the case of KIC 8429280, after computing an MCMC  of 12 million points, we find a strong correlation between the reference flux $F_{0}$ and the area of the second spot $a_{2}$. Therefore, we compute a linear regression between these two parameters with an angular coefficient $\tilde{m}$ and  impose the further constraint $\delta a_{2} = \tilde{m} \, \delta F_{0}$ that allows us to compute a well-mixed and convergent chain. In other cases, such as KIC 7985370, we find convergence by fixing the inclination $i$ and the colatitudes of the spots $\theta_{j}$ at the values corresponding to the minimum of the chi square. 

The above recipes prove useful when some information on DR is contained in the light curve, but cannot produce a well-mixed and convergent chain when this is not the case, as we found for the Sun, HD 52265, and HD 181906. If the light modulation does not provide unambiguous information to fix the longitudes of the spots and their drift vs. the time, no DR determination is possible. The MCMC algorithm makes this evident by producing a chain that is neither well-mixed nor convergent according to  Gelman and Rubin's test. In this case, by increasing the length of the chain, the values of $R$ do not approach unity and, in several cases, increase as the chain wanders among multiple separated minima in the parameter space or move along multi-dimensional correlation domains among the parameters. General methods designed to treat correlations among parameters, such as the so-called Hamiltonian MCMC \citep{Neal93,MacKay03} are  very limited or no advantage in our case, as we found by making some experiments with our stars. This happens because our problem has ten free parameters and the correlations involve several of them at the same time, often in a highly non-linear way, while the Hamiltonian method works well for low-dimensional parameter spaces where the cost of   computing  partial derivatives of the chi square vs. the parameters at each step is not high. For the same reasons,  approaches based on principal component analysis do not lead to any decisive improvement in our case, as found by \citet{Ford06} in the case of planetary transit modelling. 

\section{Model parameters}
\label{model_parameters}

The inclination of the stellar spin axis,  effective temperature,  gravity, and  limb-darkening  coefficients in the CoRoT and Kepler bandpasses as derived from the tabulation by  \citet{ClaretBloemen11}, are listed in Table~\ref{table1},  together with the appropriate  references for our stars. For $\epsilon$~Eri, we adopt a linear limb-darkening law and a spot contrast of $0.22$ in the MOST passband  to allow  a straightforward comparison of the present results with those of \citet{Crolletal06}. For Kepler-30, we adopt the quadratic limb-darkening coefficients derived by fitting  planetary transits  in the Kepler passband \citep{SanchisOjedaetal12}. The spot contrast  in all the cases, except for $\epsilon$~Eri, is fixed at the solar value, i.e, $c_{\rm s} = 0.677$ because we have no information on the starspot temperature  in our stars. The inclination has been estimated by fitting the rotational splitting of  p-mode oscillations in the case of HD~52265 and HD~181906 and has an accuracy of $\approx 10^{\circ}-15^{\circ}$. For Kepler-30 we assume that the measured projected spin-orbit obliquity of $4^{\circ} \pm 10^{\circ}$ provides evidence of an inclination of $90^{\circ}$. Also in the case of CoRoT-6, we assume alignment between the stellar spin and the orbital angular momentum of the transiting planet \citep[cf. ][]{Lanzaetal11}. For the other stars, we rely on previous multi-spot modelling with an MCMC approach to estimate the inclination  \citep{Croll06,Frascaetal11,Froehlichetal12}. This is possible with a precision up to $\pm\, 10^{\circ}-20^{\circ}$ from high-precision light curves, provided that a simple spot geometry is adopted  
\citep[cf. ][]{Cowanetal13}.
\begin{table*}
\caption{Parameters adopted in the spot modelling of the considered stars.}
\begin{tabular}{cccccccc}
 & & & & & & & \\
 \hline
Name & $i$ & $T_{\rm eff}$ & $\log \, g$ & $a_{\rm p}$ & $b_{\rm p}$ & $c_{\rm p}$ & Reference\\
 & (deg) & (K) & (cm s$^{-2}$) &  & & & \\
\hline
 & & & & & & & \\
Sun & 90 & 5777 & 4.44 & 0.360 & 0.840 & $-0.200$ & \citet{Lanzaetal03} \\
$\epsilon$ Eridani & 30 & $5100 \pm 150$ & $4.6 \pm 0.1 $ & 0.189 & 0.811 & 0.0 & \citet{Crolletal06} \\
HD 52265 & 30 & $6100 \pm 60$ & $4.35 \pm 0.1$ & 0.357 & 0.842 & $-0.199$ & \citet{Ballotetal11} \\
HD 181906 & 24 & $6300 \pm 150$ & $4.22 \pm 0.06$ & 0.368 & 0.838 & $-0.206$ & \citet{Garciaetal09} \\
CoRoT-6 & 89 & $6100 \pm 80$ & $4.44 \pm 0.1$ & 0.340 & 1.024 & $-0.378$ &  \citet{Lanzaetal11} \\
KIC 7765135 & 75 & $5835 \pm 100 $ & $4.34 \pm 0.12$ & 0.530 & 0.605 & $-0.135$ & \citet{Froehlichetal12} \\
KIC 7985370 & 41.4 & $5150 \pm 100$ & $4.24 \pm 0.12$ & 0.563 & 0.614 & $-0.141$ & \citet{Froehlichetal12} \\
KIC 8429280 & 69.5 & $5030 \pm 140$ & $4.45 \pm 0.15$ & 0.296 & 0.868 & $-0.163$ & \citet{Frascaetal11} \\
Kepler-30 & 90 & $5500 \pm 60$ & $4.7 \pm 1.0$ & 0.220 & 1.180 & $-0.400$ & \citet{SanchisOjedaetal12} \\
& & & & & & & \\
\hline 
\end{tabular}
\label{table1}
\end{table*}

\section{Results}
\label{results}
\subsection{Autocorrelation analysis}
\label{autocorr_analysis}
\begin{figure}
\centering{
\includegraphics[width=8cm,height=14cm,angle=0]{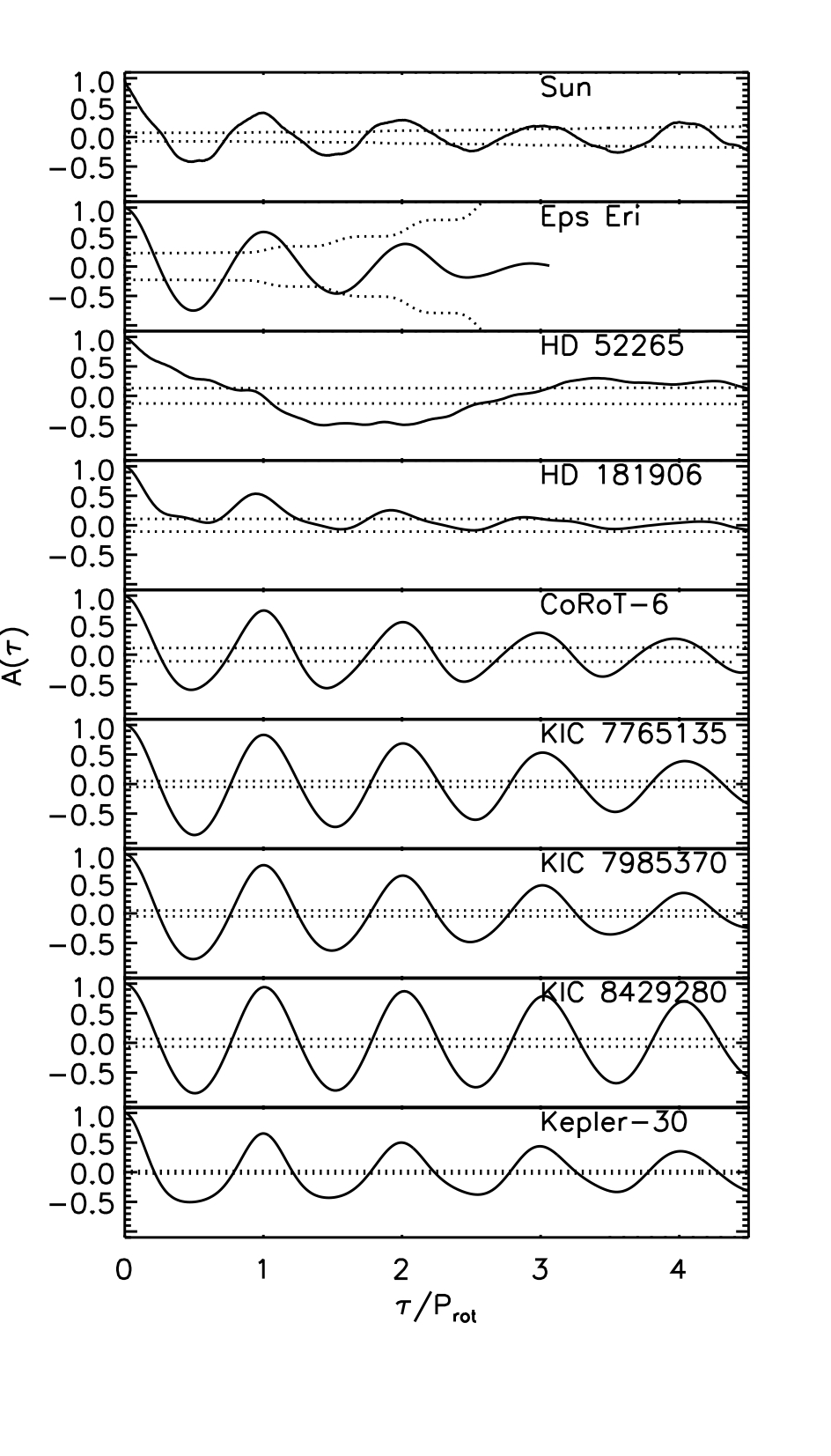}} 
\caption{Autocorrelation functions of the light curves of the stars considered in our sample. The dotted lines indicate the $\pm \sigma$ interval (see the text).}
\label{autocorr}
\end{figure}

The autocorrelation functions of the light curves of our sample stars are plotted in Fig.~\ref{autocorr} vs. the time lag measured in units of the rotation period as determined from the first maximum following the one at zero lag. The plotted lag intervals cover only a few rotation periods to show the decay of the autocorrelation over  intervals comparable to the typical duration $\Delta T$ of the intervals to which the MCMC analysis is applied to extract the DR signal.  For $\epsilon$ Eri, we have a light curve covering only three rotation periods, therefore the autocorrelation cannot be computed for longer time lags. The dotted lines indicate the interval corresponding to $\pm \sigma$, where $\sigma$ is one standard deviation of the autocorrelation as expected for a pure random noise with some degree of autocorrelation according to the large-lag approximation (see Sect.~\ref{autocorrelation}). 

For the Sun, HD~52265, and HD~181906, the secondary peaks in the autocorrelation functions are lower than $\sim 0.5$, indicating a spot-evolution timescale  that is shorter than the rotation period. For HD~52265, the evolution of starspots is remarkably faster than its rotation period ($\sim 12$ days), and the first maximum after that at  zero lag is barely visible. For the Sun and HD~181906, we can see  secondary maxima indicating some degree of coherence in the pattern of surface inhomogeneities with a timescale up to two to three rotation periods. This corresponds to $50-80$ days in the Sun as expected in the case of faculae that dominate the flux modulation near the minimum of the eleven-year cycle. However, since solar active regions are confined to $\pm 35^{\circ}$ in latitude, the expected difference in their rotation periods is only  a few percent, making it difficult or impossible to detect the DR signal since their lifetimes are much shorter than $1/\Delta \Omega$. 

\begin{figure}
\centering{
\includegraphics[width=6cm,height=9cm,angle=90]{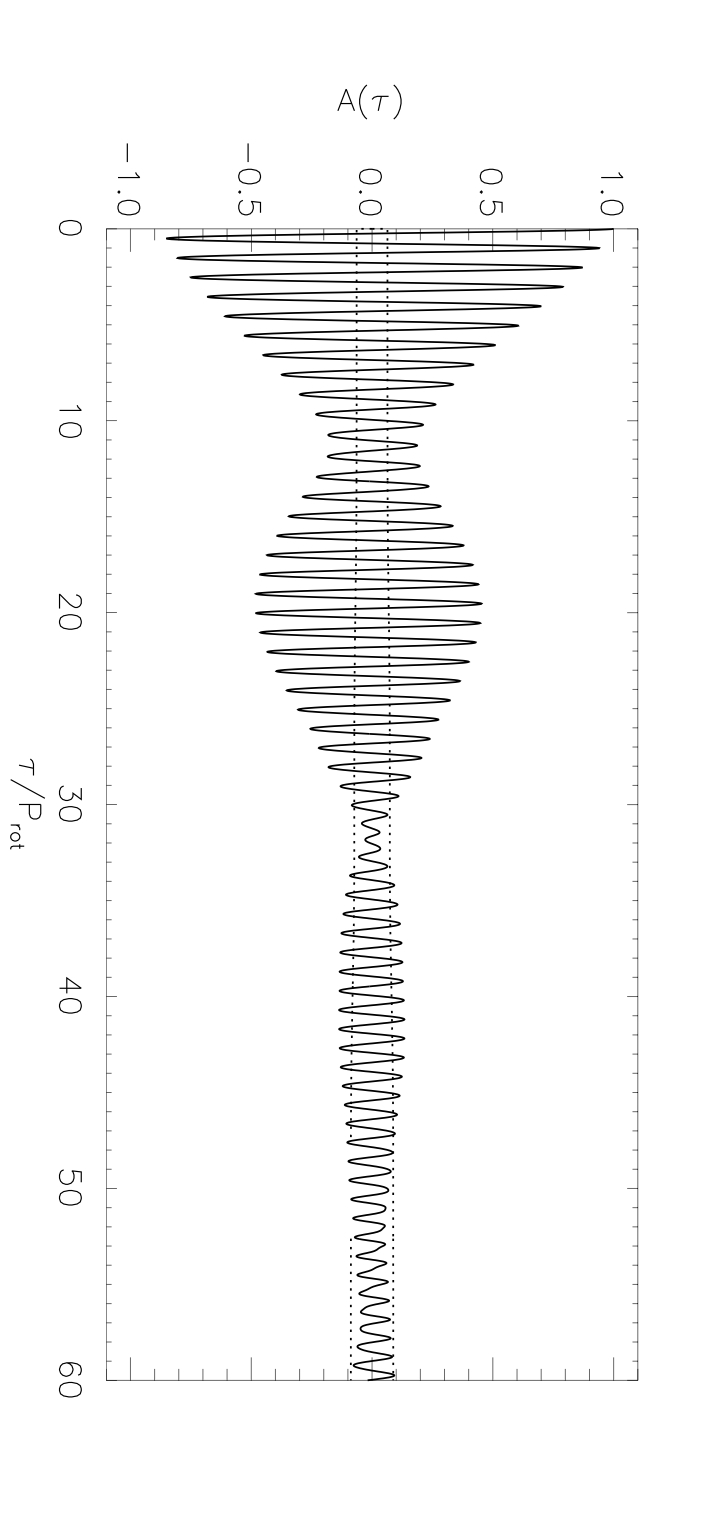}} 
\caption{Autocorrelation function of the light curve of KIC 8429280 extended up to $\tau/P_{\rm rot} = 60$ to show the beatings of the amplitude. The dotted lines indicate the $\pm \sigma$ interval  (see the text).}
\label{auto_kic8429280}
\end{figure}

One star in our sample that possibly has spots lasting longer than its rotation period is KIC~8429280, as shown by the beating modulation of the autocorrelation with an approximate half period of $\sim 20 \, P_{\rm rot}$ (see Fig.~\ref{auto_kic8429280}). In this case, we can apply the results of Sect.~\ref{autocorrelation} and  estimate a relative amplitude of DR of $\Delta P/P \approx 0.05$. Nevertheless, this value is approximate because it is based on just one beating half period, and the autocorrelation decreases rapidly reaching almost the noise level for $\tau/P_{\rm rot} \sim 10$ that suggests a strong impact of the spot evolution on this estimate. 


\subsection{The illustrative cases of Kepler-30 and $\epsilon$ Eridani}

We present detailed results for Kepler-30 and $\epsilon$~Eri because they are the star with the longest time series and the one chosen by \citet{Croll06} to develop the MCMC approach that we use with minor modifications in our analysis, respectively. Specifically, we illustrate the selection of the best  interval in the case of Kepler-30 and show the distribution of the residuals obtained with the two-spot model best fit. 
On the other hand, $\epsilon$~Eri is considered for comparing  the results of the MCMC analysis to establish the effects of the modifications we introduced in Croll's method.

\subsubsection{Kepler-30: Selection of the optimal interval  and distribution of the residuals}
\label{subinterval-selection}

For this star we computed best fits with our two-spot model for different lengths of the intervals with $M$ ranging from 9 to 65. The best case in terms of convergence of the best fits with DR and RR and time extension is found for $M=51$. For this interval (cf. Table~\ref{table2cc}), $\Delta {\rm BIC} = 6.42$ in favour of the model with DR. 

\begin{figure}
\centering{
\includegraphics[width=8cm,height=10cm,angle=90]{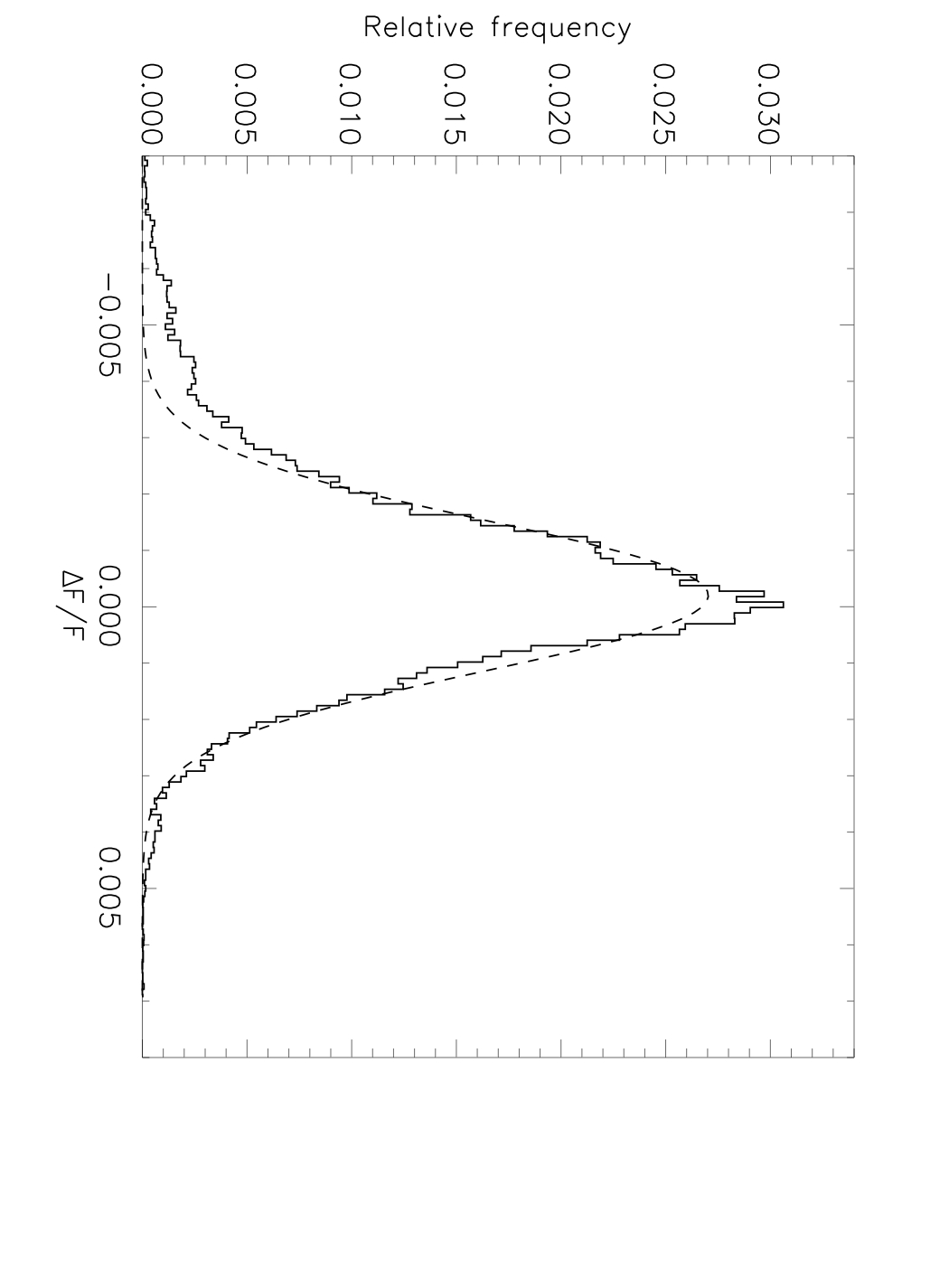}} 
\caption{Distribution of the residuals of the composite best fit to the entire time series of Kepler-30 obtained with the two-spot model with DR for $M=51$. The dashed line is a Gaussian best fit to the distribution. }
\label{res_distr}
\end{figure}

Considering a subdivision with $M=51$, which corresponds to individual time intervals $\Delta T = 22.35$ days or $1.35$ mean rotation periods, we compute the best fits with DR to the entire time series.  The distribution of their residuals is plotted in Fig.~\ref{res_distr}, together with a Gaussian fit that has a central value of $-1.97 \times 10^{-4}$ and a standard deviation of $1.33 \times 10^{-3}$ in relative flux units. 
For comparison, the precision of Kepler photometry is $\sim 2.6 \times 10^{-4}$, i.e., about five times less than the standard deviation of the residuals implying that small unaccounted spots (or possibly flares) play a relevant role in determining the residuals of our model. The distribution shows significant deviations for negative values below $-2 \times 10^{-3}$, suggesting that  dark spots are responsible for the largest residuals.  The interesting point is that the  distribution of the residuals does not strongly deviate from a Gaussian one, thus supporting our application of the BIC approach for a preliminary selection of the intervals containing some DR signal.

\subsubsection{Two-spot modelling of $\epsilon$ Eridani light curve}

\begin{figure}
\centering{
\includegraphics[width=8cm,height=10cm,angle=90]{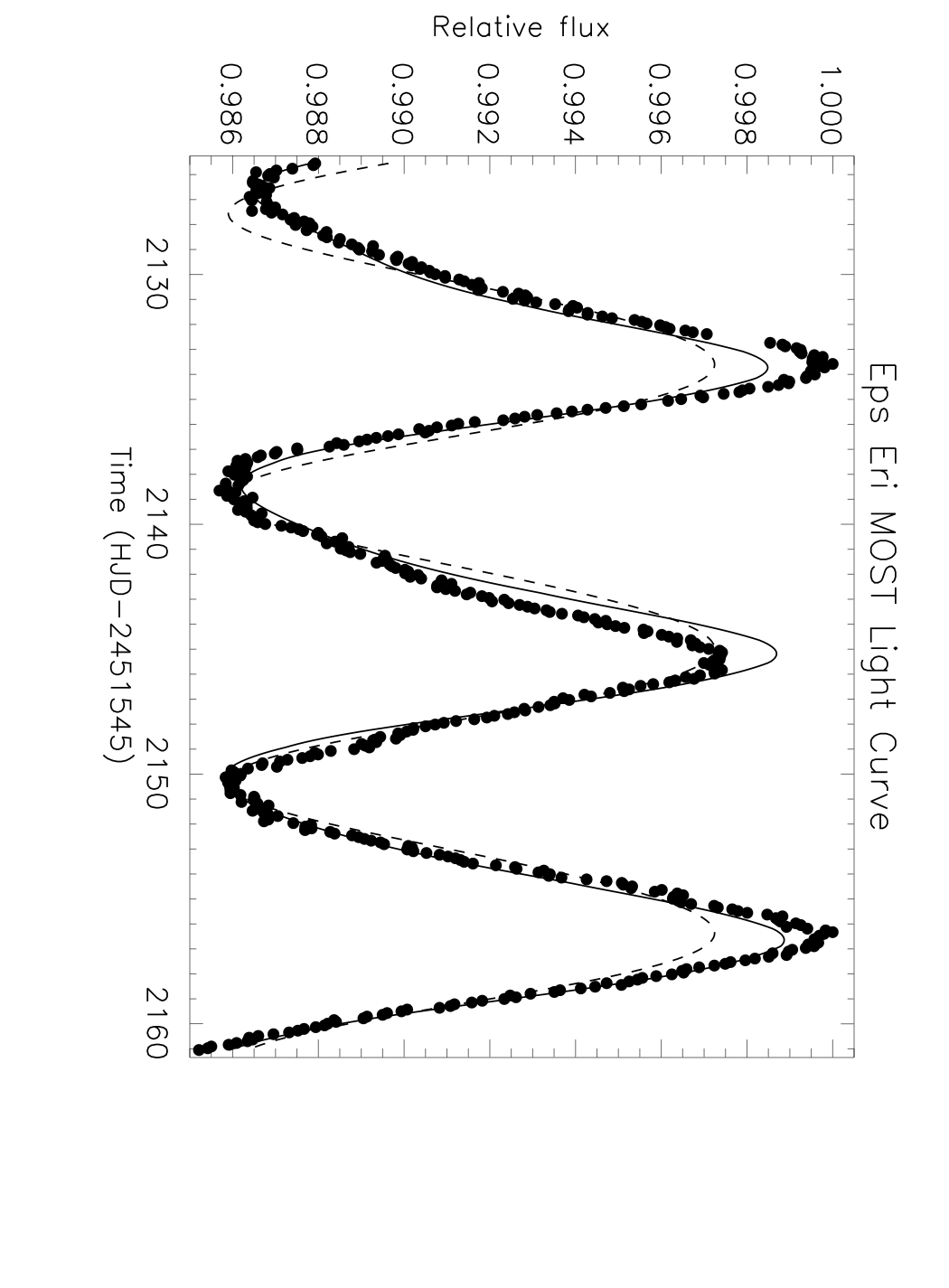}} 
\caption{The optical flux of $\epsilon$ Eridani (filled dots)  vs. the time, together with the best fits obtained with our two-spot model with DR (solid line) or  rigid rotation (dashed line).}
\label{eps_eri_two_spots}
\end{figure}

The optical light curve of $\epsilon$~Eri, together with the best fits obtained with our model, is plotted in Fig.~\ref{eps_eri_two_spots}. 
The best fit with DR gives a better reproduction of the variable amplitude of the light modulation along the nearly three rotations covered by the dataset, while the best fit with two rigidly rotating spots has larger deviations and does not reproduce the times of all three light minima. In particular, the second and the third minima are approximately matched, while the first one is missed because the two deeper minima strongly constrain the (unique) spot rotation period. On the other hand, the model with DR provides a significantly better fit because the two spots can drift in longitude with respect to each other, thus allowing a simultaneous fit of all the three minima.  The value of $\Delta {\rm BIC}$ computed according to Eqs.~(\ref{delta_bic}) and~(\ref{delta_likely}) is $1.94$ in favour of the model with DR. This difference is close to the threshold adopted for selecting cases with some evidence of  DR so it gives support to our choice. 

\subsubsection{MCMC analysis for $\epsilon$ Eridani and comparison with previous results}

We computed an MCMC chain of 48 million points starting from the minimum  $\chi^{2}$ found with the two-spot model with DR  as given by the Levenberg-Marquardt algorithm. The standard deviations of the random steps of the parameters were chosen as to obtain an overall acceptance rate of 0.27. 
The  prior distributions of our parameters were the uniform wide priors  in Table~1 of \citet{Croll06} with the exception of the  reference flux level $F_{0}$  that we assumed to range from $-0.001$ and $0.005$ and the use of initial spot longitudes instead of  spot transiting epochs in our model; we assumed a range of $\pm 13^{\circ}$ for the initial longitudes. 
The mixing and convergence of the chain were checked as explained in Sect.~\ref{mcmc_method}, obtaining a minimum value of the parameter $R$ of 1.0017 and a maximum value of 1.0133 { (see Table~\ref{table_rvalues})}. Therefore, the convergence of our chain for $\epsilon$~Eri was remarkably good. In the present approach, we considered point-like active regions, while \citet{Croll06} assumed circular (polar cap) spots. Moreover, we limited the maximum $\chi^{2}$ variation for the acceptance of a given step to one percent, which was significantly less than that of Croll who adopted a limit around four percent \citep{Crolletal06}. Croll's choice of the odds relating one point to the next in the chain was ruled by a virtual temperature according to his Eq.~(1). We assumed a similar exponential dependence,  but dropped the factor 2 present in the denominator of his equation (cf. Sect.~\ref{mcmc_method}). 
In spite of all these differences, our results were very similar to Croll's. 

\begin{figure}
\centering{
\includegraphics[width=8cm,height=15cm,angle=0]{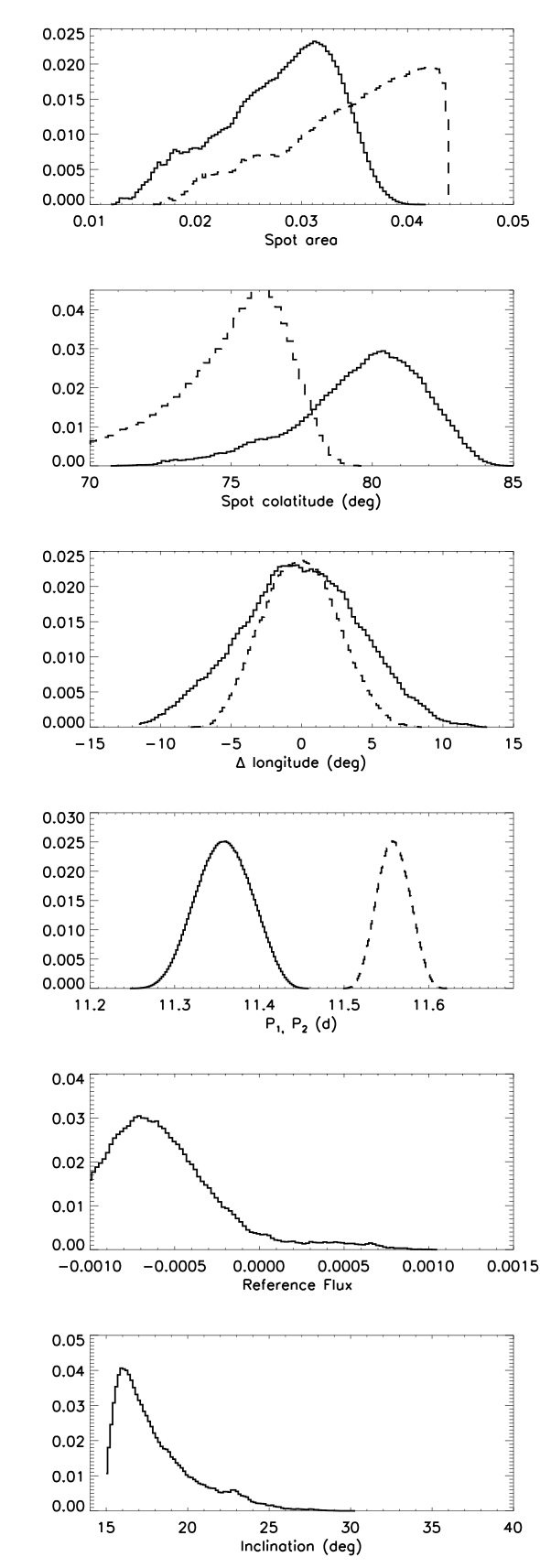}} 
\caption{Frequency marginal distributions of the parameters of the two-spot model with DR for $\epsilon$ Eridani. From top to bottom: area of the first (solid line) and the second (dashed line) spot; colatitude of the first (solid line) and the second (dashed line) spot; deviation of the longitude of the first spot (solid line) and of the second spot (dashed line) from their initial longitudes, respectively;  rotation period of the first spot (solid line) and of the second spot (dashed line);  reference flux;  inclination of the stellar spin axis.}
\label{mcmc_distr}
\end{figure}

The marginal distributions of the model parameters, as derived a posteriori from our chain of 48 million points to which a thinning factor of 10 has been applied, are shown in Fig.~\ref{mcmc_distr}. One can compare these distributions with those plotted in Fig.~3 of \citet{Croll06}, considering that we plot the distributions of the spot areas and not of the radii. Moreover, we do not consider the mean likelihood distributions of the parameters. Our spot colatitudes are not significantly different from Croll's spot latitudes, while our areas are in general agreement with the values of his radii, although our assumption of point-like spots makes the flux modulation induced by a given spot slightly different. Our range of approximately $\pm 13^{\circ}$ in  spot initial longitude corresponds to a range of about $\pm 0.4$ days in spot transit epoch  and leads to a general agreement with  Croll's distributions. 
The distributions of spot rotation periods,  upon which our measurement of DR is based, 
are remarkably similar to those of Croll indicating that the above differences between the models do not significantly affect the determination of DR. Nevertheless, the distribution of the inclination is significantly different, as expected owing to the strong dependence of this parameter on details of the spot model, in particular on the adopted spot geometry and the level of unspotted flux. 

\begin{figure}
\centering{
\includegraphics[width=8cm,height=8cm,angle=90]{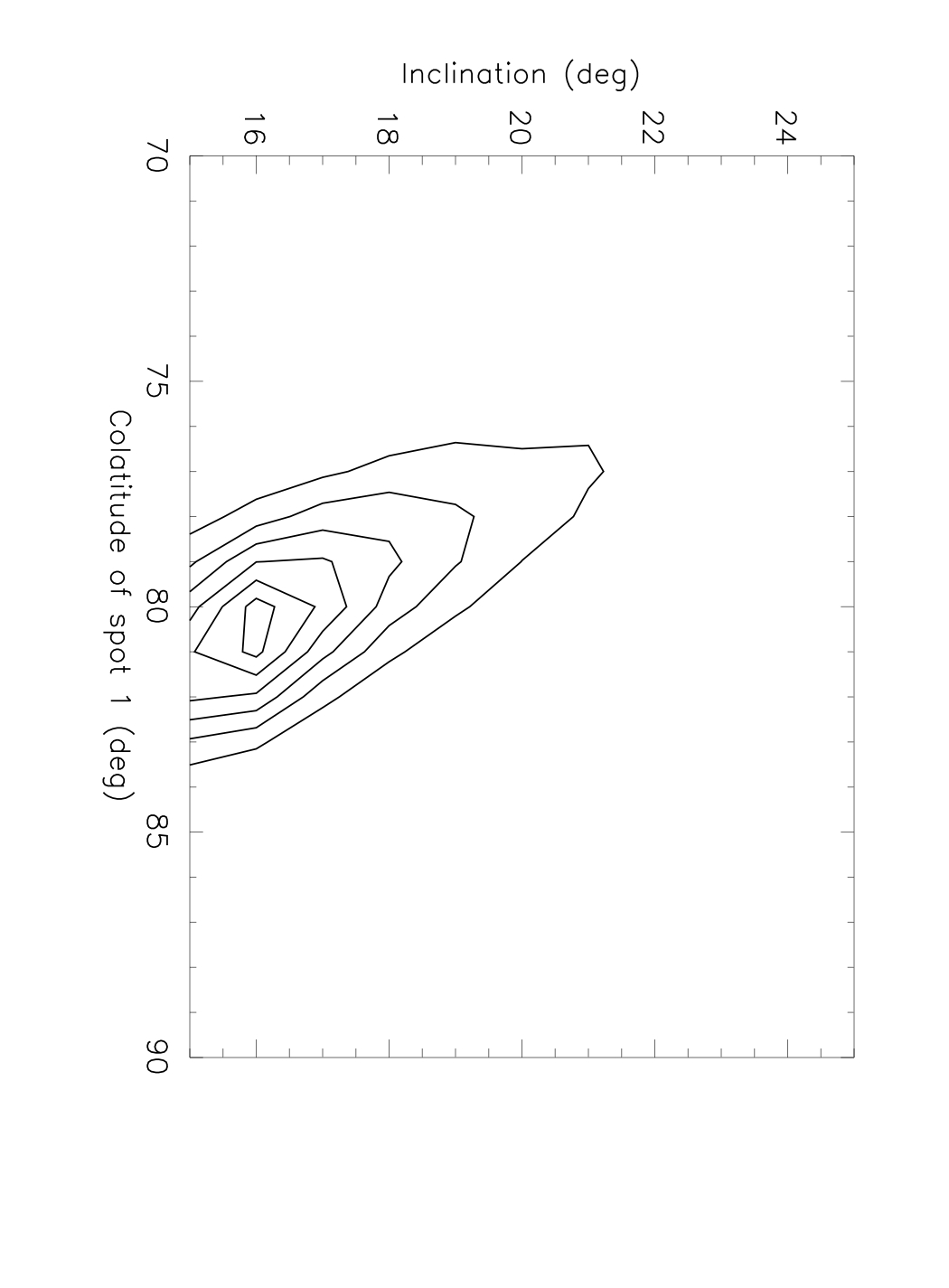}} 
\caption{Isocontour plots of the joint distribution of the inclination of the stellar rotation axis vs. the colatitude of the first spot for $\epsilon$ Eridani. }
\label{theta1_incl}
\end{figure}

\begin{figure}
\centering{
\includegraphics[width=8cm,height=8cm,angle=90]{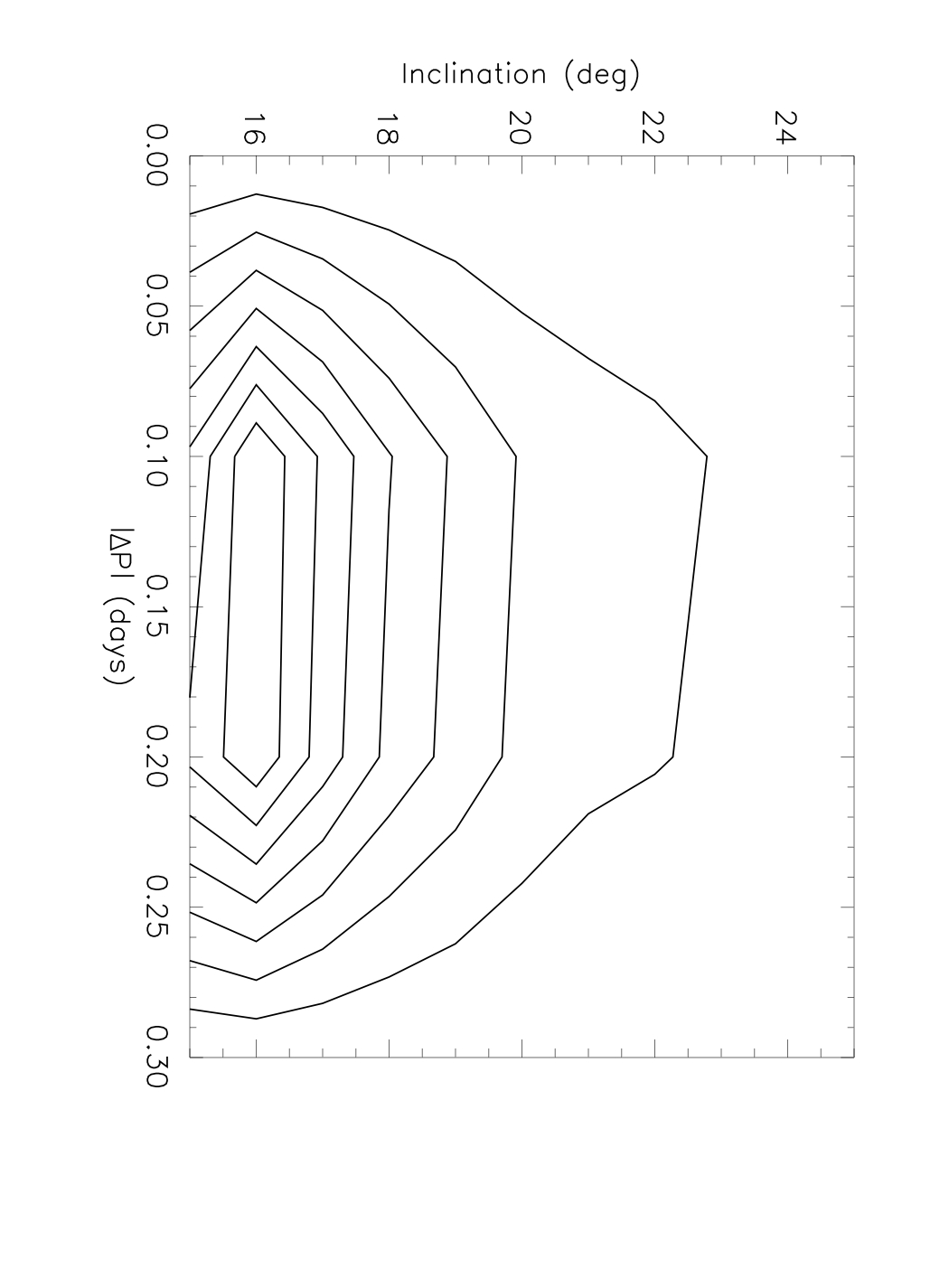}} 
\caption{Isocontour plots of the joint distribution of  the inclination of the stellar rotation axis vs. the absolute value of the difference between the spot rotation periods  for $\epsilon$ Eridani. }
\label{deltap_incl}
\end{figure}

We find similar correlations among the model parameters as in Fig.~2 of \citet{Croll06}. For instance, we plot the correlation between the colatitude of the first spot and the inclination of the stellar rotation axis in Fig.~\ref{theta1_incl}. This correlation produces a remarkable dependence of the parameter $k$ of Croll, measuring the relative amplitude of the pole-equator DR, on the inclination (see his Fig.~2, lower right panel). In our case, we adopt the absolute value of the difference between the spot rotation periods $\Delta P \equiv | P_{1} -P_{2}|$ as a measure of the DR. Unlike $k$, it does not depend on the spot colatitudes as derived from the modelling and is therefore a robust quantity. This is confirmed by the plot in Fig.~\ref{deltap_incl} that shows that the distribution of $\Delta P$ does not especially depend on $i$, thus allowing  a robust estimation of the relative amplitude of the DR as $\Delta P /P$, where $P = (P_{1} + P_{2})/2$. A similar conclusion has been  reached by \citet{Froehlich07}, who also applied a Bayesian spot modelling with different priors.


\subsection{DR in our sample of stars}
\label{DR_results}

For the  eight stars in our sample, including the Sun, we  determine the best intervals for measuring the DR according to the method introduced in Sect.~\ref{interval_selection} and illustrated by the case of Kepler-30 in Sect.~\ref{subinterval-selection}.   The beginning and the end of these intervals are indicated for each star in Table~\ref{table2cc}, together with the uniform priors applied to the model parameters. The origin of the time is in all  cases the first observation of the corresponding time series (cf. Sect.~\ref{observations}), while the other symbols are defined in Sect.~\ref{spot_modelling}. The longitude intervals are defined as the maximum allowed deviations from the initial point of the MCMC chain that correspond to the parameters giving the minimum  $\chi_{\rm DR}^{2}$ (see below for details).  

For the Sun, HD~52265, and HD~181906, the MCMC chains do not properly converge as indicated by  values of the  parameter $R > 1.2$ for one or several of the model parameters { (see Table~\ref{table_rvalues})}. No significant improvement is obtained by applying the recipes described in Sect.~\ref{param_corr} or by reducing the acceptance threshold for the chi square variation to values as low as $0.1-0.5$ percent.   For these stars, with the exception of the Sun, we repeat the analyses with a different interval to confirm that the result does not depend on the particular  interval (cf. Table~\ref{table2cc}). { The values of  $R$  given in Table~\ref{table_rvalues} provide some information on the model parameter chains that have not reached a proper convergence or mixing. We note that  $R$ is undefined when the corresponding model parameter has been fixed in an attempt to improve the convergence as in the case of the Sun \citep[cf. Sect.~3.2 in ][]{Verdeetal03}. Values of $R$ lower than the acceptable threshold of $1.2$ for only a subset of the model parameters do not imply that the marginal distributions of the corresponding parameters have converged,  because the correlations with the parameters having $R> 1.2$ make them unreliable. Therefore, we require that $R< 1.2$ for all the parameters in order to use the a posteriori distributions derived with the MCMC approach for the individual parameters. 
}
It is interesting to note that all these stars for which MCMCs do not convergence are characterized by an autocorrelation function with secondary peaks that decrease rapidly with increasing time lag, thereby indicating a fast spot evolution. Previous analysis based on Lomb-scargle periodogram and multi-spot models have provided some evidence of DR in those stars, although they also detected a rapid intrinsic spot evolution that could  affect the determination, possibly mimicking a DR signal \citep{Ballotetal11,Mosseretal09}. 

For CoRoT-6, KIC~7765135, KIC~7985370, KIC~8429280, and Kepler-30, we are able to compute convergent and well-mixed MCMCs by applying the recipes in Sect.~\ref{param_corr} and fixing the threshold for the chi square acceptance between one and five percent above the minimum {(cf. Table~\ref{table_rvalues})}.  For Kepler-30 and KIC~7765135, we find  maximum values of $R$ of 1.197 and 1.067 considering chains of 36 and 500 million steps, respectively. In the case of CoRoT-6, the maximum value of $R$ is 1.186  considering a chain of 128 million steps. For KIC~7985370, it has  been necessary to fix the inclination of the rotation axis and the colatitudes of the starspots at the values corresponding to the minimum of the chi square to obtain a maximum $R$ equal to 1.125 for a chain of 180 million steps. { Therefore, the $R$ values corresponding to those fixed parameters are undefined. }
For KIC~8429280, we find strong correlations among the reference flux $F_{0}$, the inclination $i$, and the spot area $a_{2}$ that can be treated with the approach described in Sect.~\ref{param_corr}, which yields a maximum $R$ of 1.010 after computing a chain of 36 million steps. { As a consequence of the correlations explicitly included in the model, the values of $R$ corresponding to those parameters are the same in Table~\ref{table_rvalues}.}

\begin{figure}
\centering{
\includegraphics[width=8cm,height=8cm,angle=90]{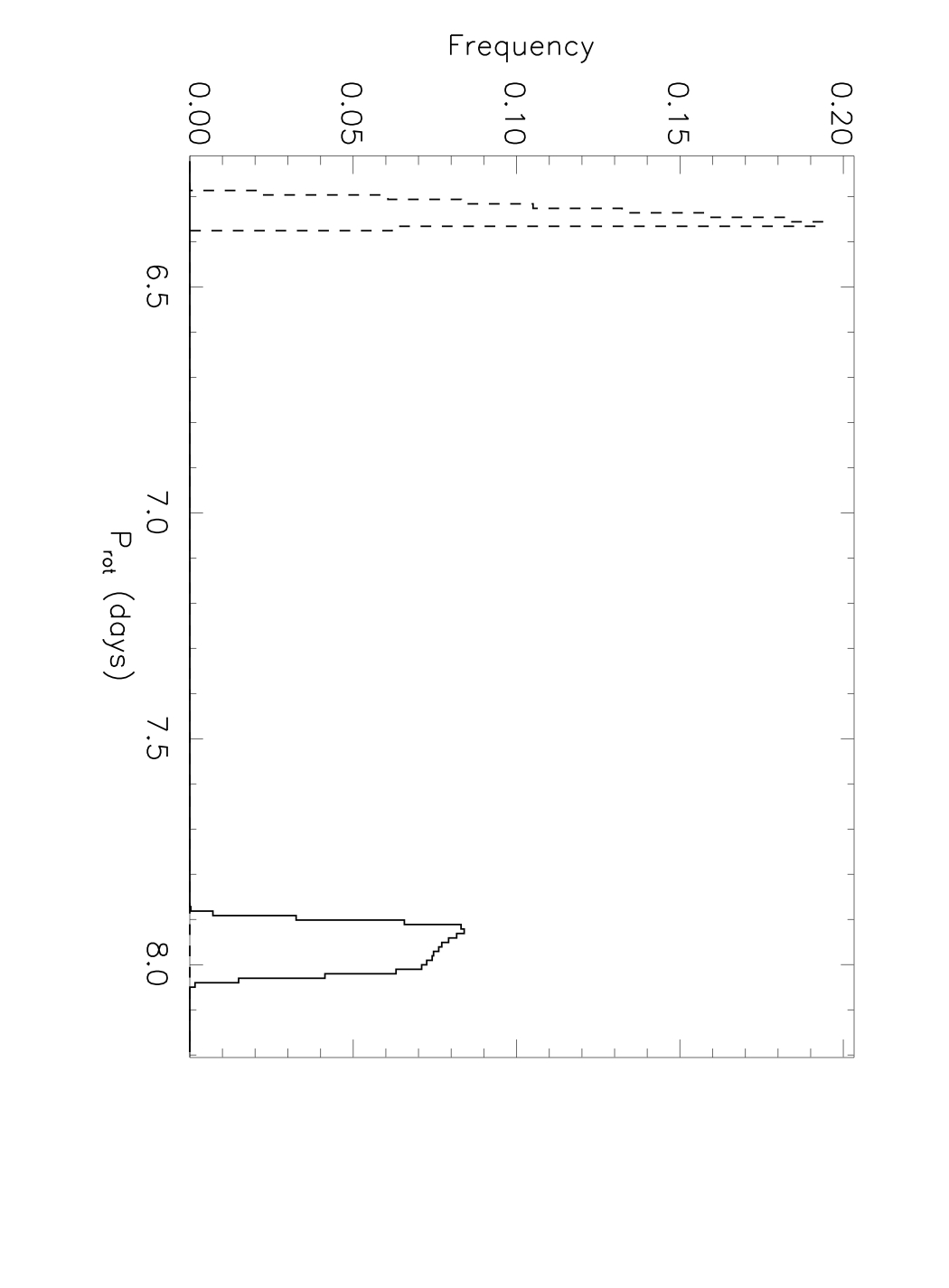}} 
\caption{A posteriori distributions of the rotation periods of the two spots as derived from  MCMC analysis for CoRoT-6. The solid line refers to the distribution of the rotation period of the first spot, the dashed line to that of the second.}
\label{prot_corot6}
\end{figure}

\begin{figure}
\centering{
\includegraphics[width=8cm,height=8cm,angle=90]{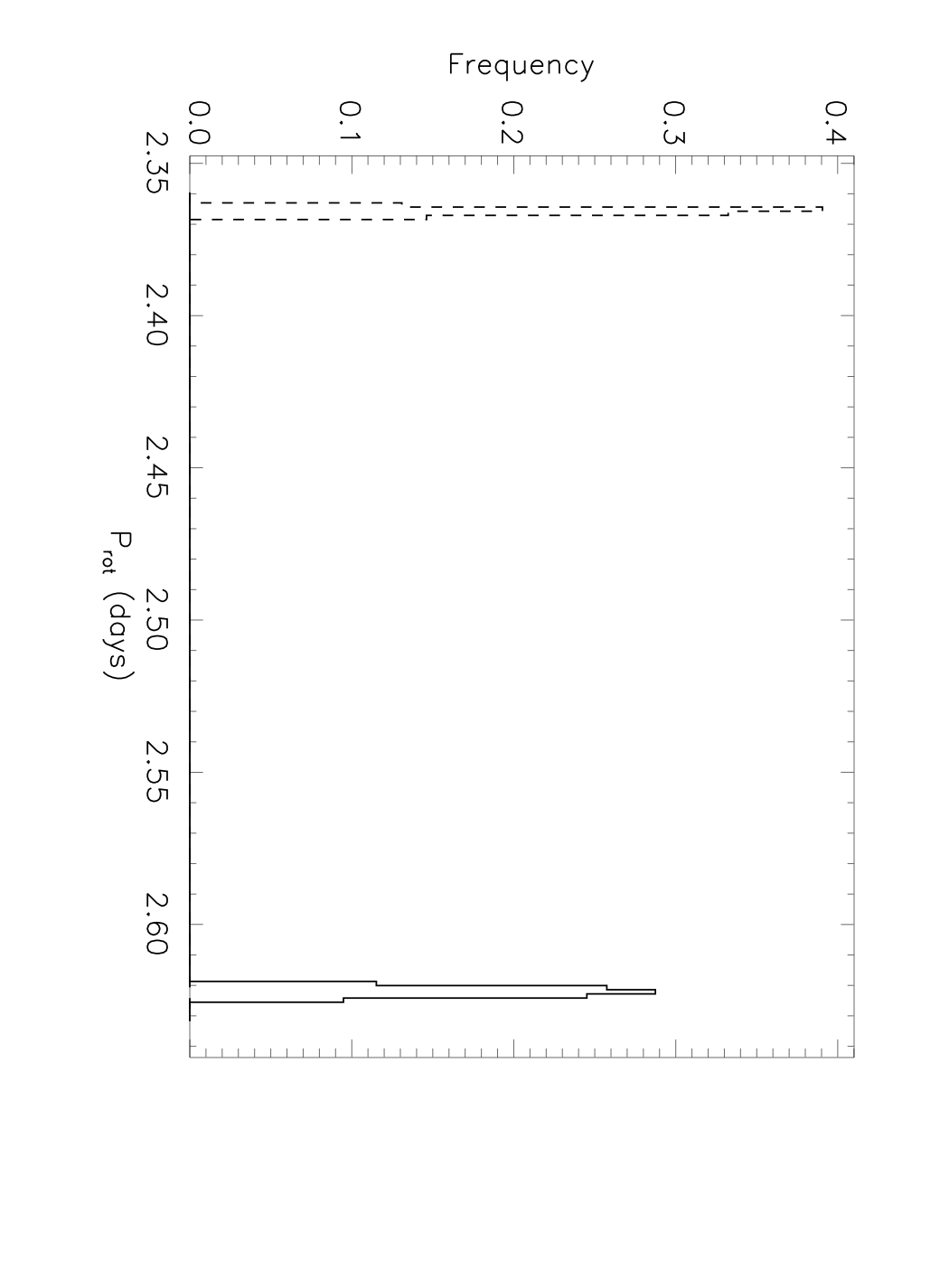}} 
\caption{Same as Fig.~\ref{prot_corot6} for KIC~7765135.}
\label{prot_kic7765135}
\end{figure}

\begin{figure}
\centering{
\includegraphics[width=8cm,height=8cm,angle=90]{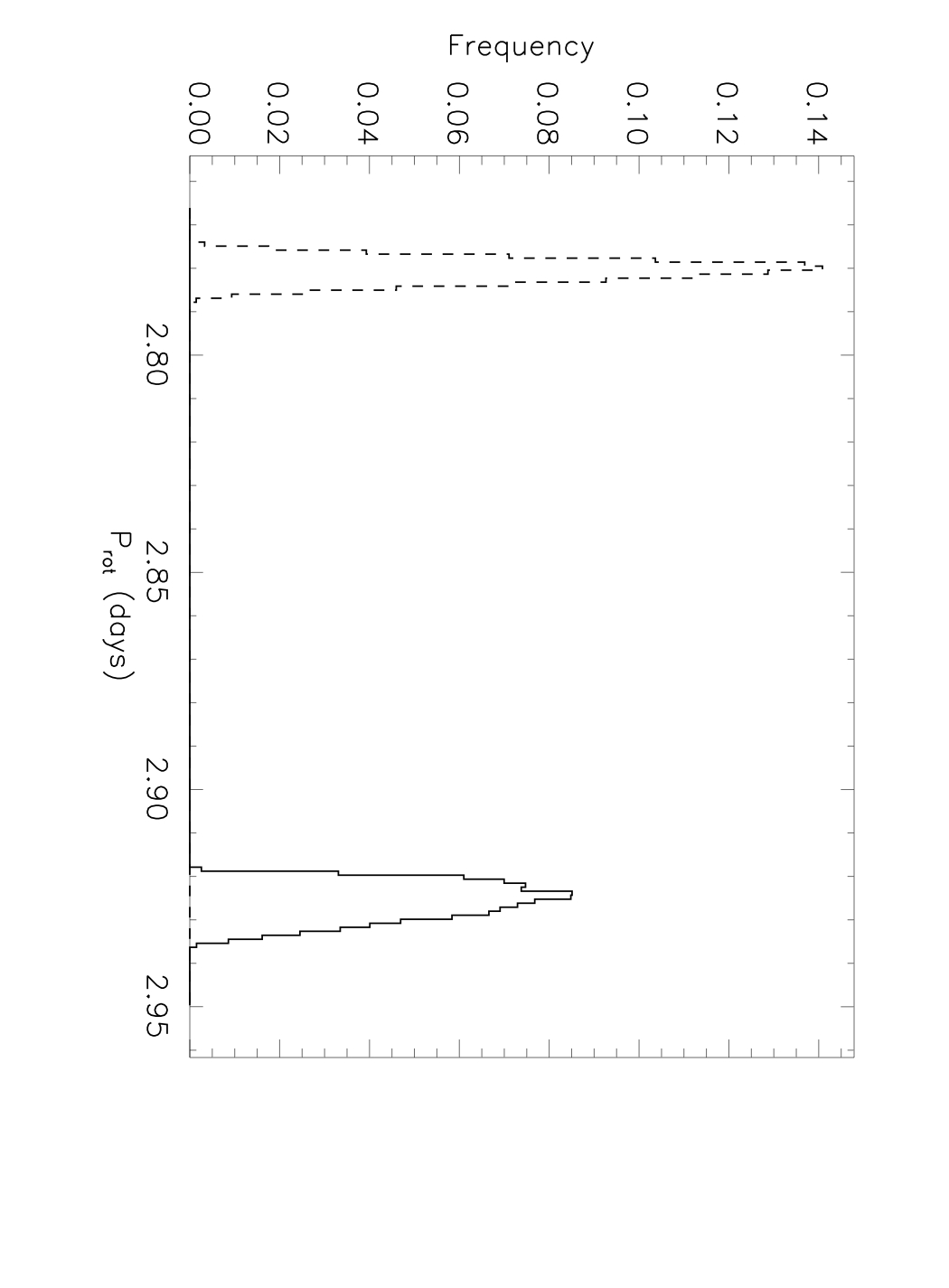}} 
\caption{Same as Fig.~\ref{prot_corot6} for KIC~7985370.}
\label{prot_kic7985370}
\end{figure}

\begin{figure}
\centering{
\includegraphics[width=8cm,height=8cm,angle=90]{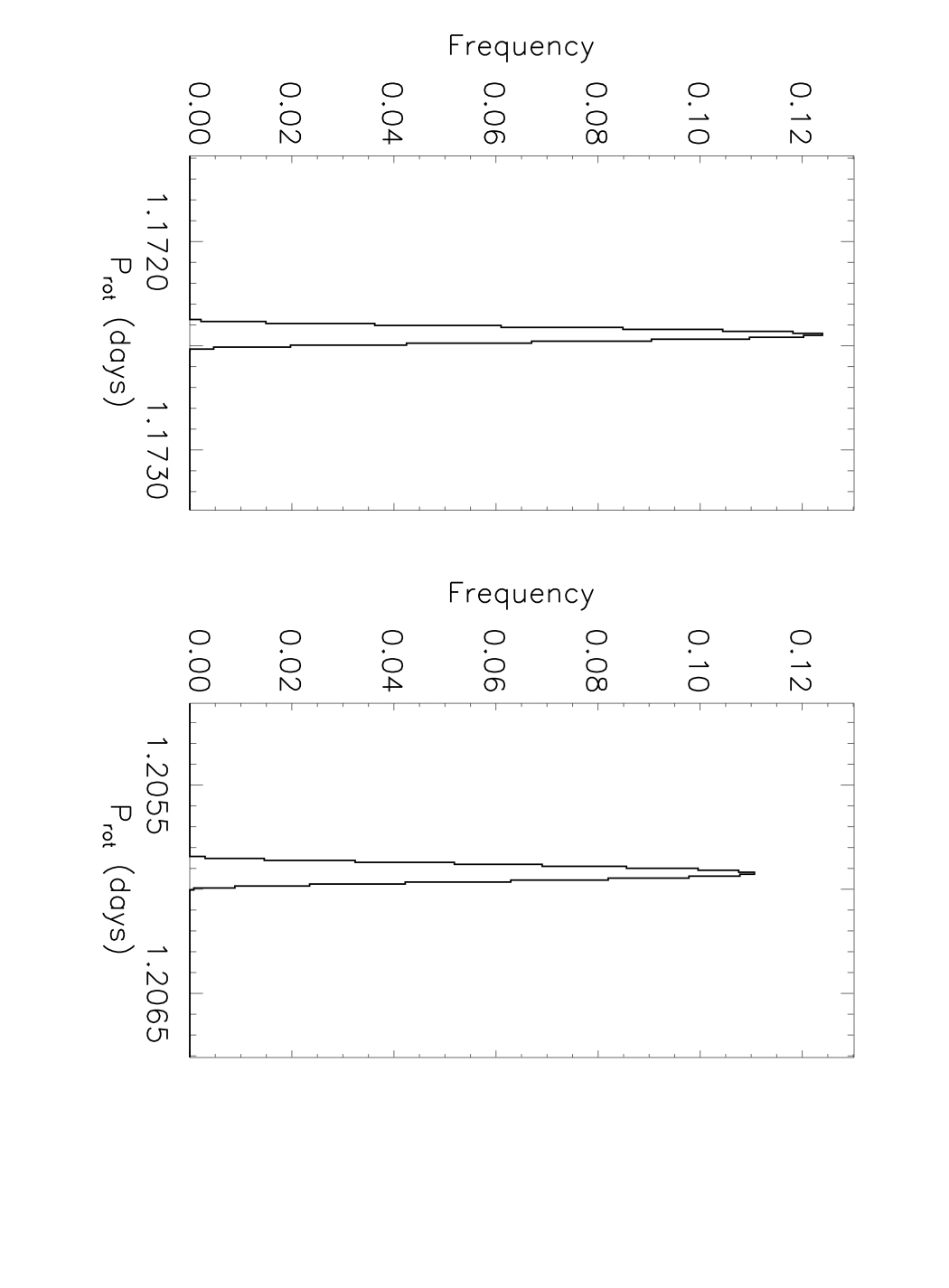}} 
\caption{Same as Fig.~\ref{prot_corot6} for the distributions of the rotation periods of the two spots of KIC~8429280.}
\label{prot_kic8429280}
\end{figure}

\begin{figure}
\centering{
\includegraphics[width=8cm,height=8cm,angle=90]{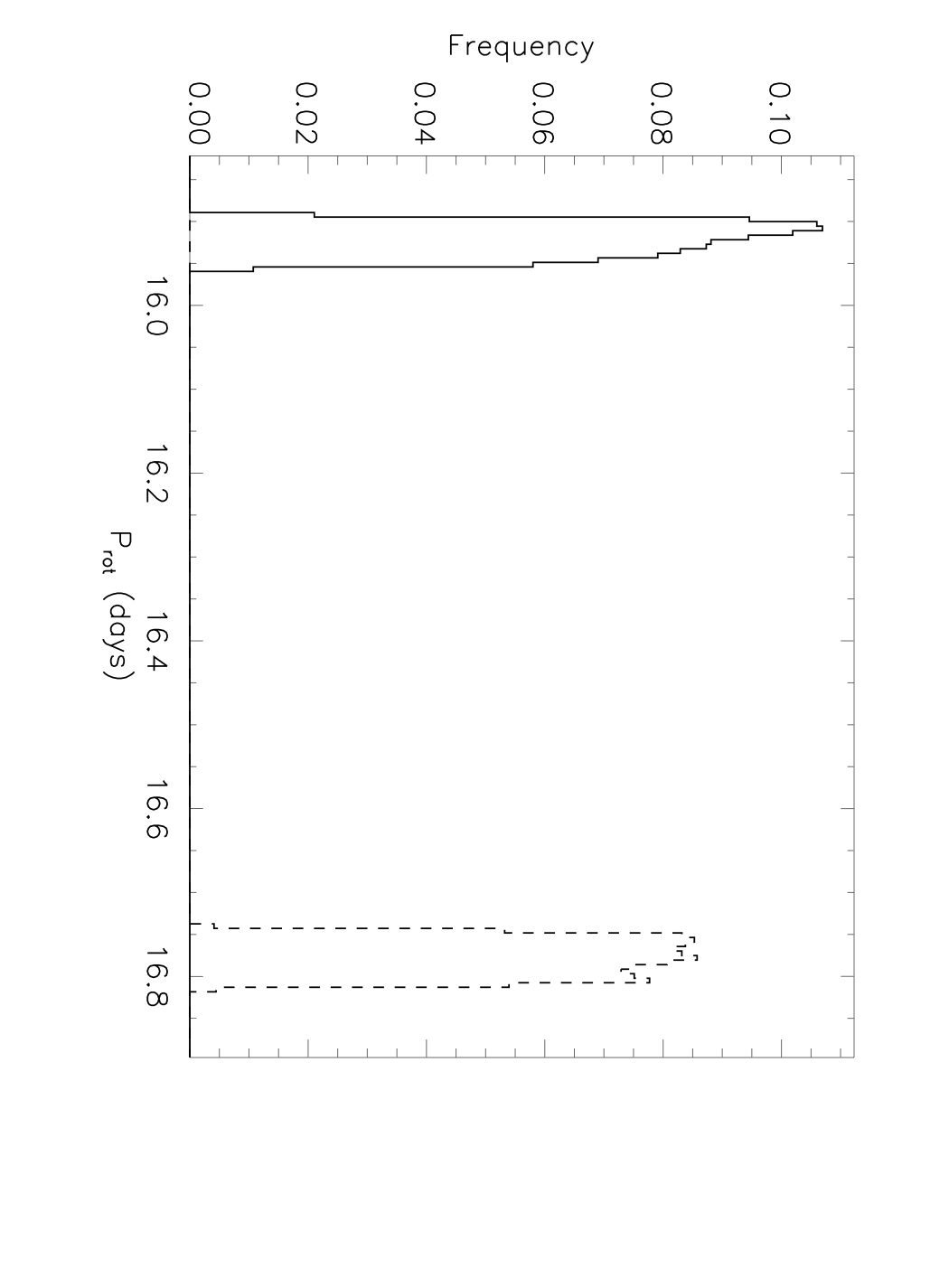}} 
\caption{The same as Fig.~\ref{prot_corot6} for Kepler-30.}
\label{prot_kepler30}
\end{figure}

In Table~\ref{table3}, we list the rotation periods of the two starspots with their standard deviations as derived from the a posteriori marginal distributions of $P_{1}$ and $P_{2}$ as well as the corresponding relative amplitude of the differential rotation $\Delta P /P$ with its standard deviation for our sample stars. 
 The distributions of the rotation periods of the two starspots are plotted in Figs.~\ref{prot_corot6}~(CoRoT-6), \ref{prot_kic7765135}~(KIC~7765135), \ref{prot_kic7985370}~(KIC~7985370), \ref{prot_kic8429280}~(KIC~8429280), and~\ref{prot_kepler30}~(Kepler-30). The area of the intersection between the distributions  is virtually zero, thus providing evidence for DR in the framework of our model. 

The amplitude of the DR determined by our method depends of course on the specific interval considered because spots can appear at different latitudes in late-type stars. The latitude range is wider in more active stars than in the Sun \citep[see, e.g., ][]{Mossetal11}, thus making the differences larger for CoRoT-6 and the young Sun-like stars in our sample. Specifically, from the modelling of another interval for CoRoT-6, we find $\Delta P/P = 0.259 \, \pm \, 0.003$, while for, say, KIC~8429280, another interval gives $\Delta P /P = 0.011\, \pm \, 3.73 \times 10^{-5}$, where the standard deviations are derived from the a posteriori distributions given by the MCMC method. This should be kept in mind when comparing the present results with previous ones.   For CoRoT-6, \citet{Lanzaetal11} find $\Delta P / P$ of about half the present value by tracing the migration of the active longitudes, not of the individual spots. A remarkable difference in those migration rates has indeed been found in CoRoT-2 that could explain the present result  \citep[cf.  ][]{Lanzaetal09}. Our spot rotation periods fall within the range found by \citet{Frascaetal11} for KIC~8429280, while they are close but slightly outside the ranges found for KIC~7765135 and KIC~7985370 by \citet{Froehlichetal12}. However, those models use seven to nine evolving spots to fit the whole time series instead of two fixed spots applied to fit a data interval. 
\begin{table*}
\begin{center}
\caption{Initial and final times of the intervals considered for the MCMC analysis, together with the  uniform prior intervals of the parameters for the stars of our sample. }
\begin{tabular}{ccccccccc}
\hline
 Star name & $t_{1}$ & $t_{2}$ & $i$ & $F_{0}$ & $a_{1,2}$ & $\theta_{1, 2}$ & $\Delta \lambda_{1,2}$ & $P_{1, 2}$ \\
  & (d) & (d) & (deg) & & & (deg) & (deg) & (d) \\
  \hline
  Sun & 19.8542 & 39.7084 & 70,90 & $-0.03$, 0.01 & $2 \times 10^{-5}$, 0.060 & 0, 180 & $\pm 13$ & 24.0, 29.0 \\
  HD~52265 & 48.3125 & 60.2654 & 15, 45 & $-2 \times 10^{-4}$, $ 2\times 10^{-4}$ & $2 \times 10^{-7}$, 0.044 & 0, 150 & $\pm 15$ & 9.0, 14.0 \\
  HD~52265 & 60.3210 & 72.3890 & 15, 45 & $-2 \times 10^{-4}$, $ 2\times 10^{-4}$ & $2 \times 10^{-7}$, 0.044 & 0, 150 & $\pm 15$ & 9.0, 14.0 \\
  HD~181906 & 23.1900 & 28.9158 & 15, 40 & $-0.001$, 0.002 & $2 \times 10^{-5}$, 0.044 & 0, 120 & $\pm 13$ & 2.65, 2.90 \\
  HD~181906 & 37.9341 & 42.6582 & 15, 40 & $-0.001$, 0.002 & $2 \times 10^{-5}$, 0.044 & 0, 120 & $\pm 13$ & 2.65, 2.90 \\
 CoRoT-6 & 75.930140 & 82.810186 & 70, 90 & $-0.03$, 0.01 & $2 \times 10^{-3}$, 0.268 & 0, 180 & $\pm 13$ & 5.0, 9.0 \\
 KIC~7765135 & 88.07162 & 93.91567 & 50, 60 & $-0.03$, 0.01 & $2 \times 10^{-3}$, 0.268 & 0, 180 & $\pm 13$ & 1.0, 4.0 \\
 KIC~7985370 & 33.5126 & 39.07074 & 25, 50 & $-0.03$, 0.01 & $2 \times 10^{-3}$, 0.268 & 0, 180 & $\pm 13$ & 1.0, 5.0 \\
 KIC~8429280 & 104.49977 & 108.66819 & 65, 73 & $-0.02$, 0.01 & $2 \times 10^{-3}$, 0.268 & 0, 180 & $\pm 13$ & 1.0, 4.0 \\
 Kepler-30 & 290.99150 & 313.34662 & 60, 90 & $-0.03$, 0.01 & $2 \times 10^{-5}$, 0.060 & 0, 180 & $\pm 13$ & 13.0, 19.0 \\
  \hline
\end{tabular}
\label{table2cc}
\end{center}
\end{table*}
\begin{table*}
\begin{tiny}
\begin{center}
\caption{The parameter $R-1$ of Gelman and Rubin measuring the convergence and mixing of the MCMC chains for the different parameters of our two-spot model.}
\begin{tabular}{ccccccccccc}
\hline
Star name & $i$ & $F_{0}$ & $a_{1}$ & $\theta_{1}$ & $\lambda_{1}$ & $P_{1}$ & $a_{2}$ & $\theta_{2}$ & $\lambda_{2}$ & $P_{2}$ \\
\hline
Sun & $-$ & $3.39 \times 10^{-2}$ & $3.03 \times 10^{-2}$ &  $-$ & 3.52 & 3.52 & $2.14 \times 10^{-2}$ & $-$ & 4.94 & 4.94 \\
$\epsilon$~Eridani & $9.34 \times 10^{-3}$ & $ 1.33 \times 10^{-2}$ & $9.32 \times 10^{-3}$ & $9.74 \times 10^{-3}$ & $6.12 \times 10^{-3}$ & $4.26 \times 10^{-3}$ & $9.66 \times 10^{-3}$ & $1.25 \times 10^{-2}$ & $6.55\times 10^{-3}$ & $1.70 \times 10^{-3}$ \\
HD~52265 & $5.26 \times 10^{-2}$ & $4.39 \times 10^{-3}$ & $2.41 \times 10^{-3}$ & 3.19 & 20.64 & $2.80 \times 10^{-3}$ & $7.31 \times 10^{-4}$ & 2.75 & 8.03 & $5.61 \times 10^{-3}$ \\
HD~52265 & 0.42 & $5.54 \times 10^{-2}$ & 0.22 & 13.12 & 4.98 & $3.49 \times 10^{-3}$ & 0.12 & 33.44 & 6.09 & $6.99 \times 10^{-2}$ \\
HD~181906 & $1.86 \times 10^{-4}$ & $7.91 \times 10^{-5}$ & $2.60 \times 10^{-5}$ & 12.91 & $4.23 \times 10^{-2}$ & $5.43 \times 10^{-5}$ & $1.10 \times 10^{-5}$ & 1.71 & $8.85 \times 10^{-2}$ & $3.86 \times 10^{-5}$ \\
HD~181906 & $4.58 \times 10^{-2}$ & $4.27 \times 10^{-2}$ & $2.96 \times 10^{-2}$ & 10.81 & 0.65 & $9.93 \times 10^{-5}$ & $3.03 \times 10^{-2}$ & 5.63 & 0.19 & $9.97 \times 10^{-4}$ \\
CoRoT-6 & $1.94 \times 10^{-2}$ & $7.72 \times 10^{-3}$ & $2.24 \times 10^{-2}$ & $2.34 \times 10^{-2}$ & $4.99 \times 10^{-2}$ & $5.13 \times 10^{-2}$ & $4.46 \times 10^{-2}$ & $4.70 \times 10^{-2}$ & 0.19 & 0.19 \\
KIC~7765135 & $2.82 \times 10^{-2}$ & $3.29 \times 10^{-2}$ & $1.14 \times 10^{-2}$ & $3.12 \times 10^{-2}$ & $6.68 \times 10^{-2}$ & $ 6.62 \times 10^{-2}$ & $1.82 \times 10^{-2}$ & $2.90 \times 10^{-2}$ & $5.60 \times 10^{-2}$ & $6.01 \times 10^{-2}$ \\
KIC~7985370 & $-$ & $6.23 \times 10^{-2}$ & $7.62 \times 10^{-2}$ & $-$ & $8.53 \times 10^{-2}$ & $9.16 \times 10^{-2}$ & $5.41 \times 10^{-2}$ & $-$ & 0.12 & 0.12 \\
KIC~8429280 & $6.62 \times 10^{-3}$ & $1.03 \times 10^{-2}$ & $3.24 \times 10^{-3}$ & $6.62 \times 10^{-3}$ & $8.06 \times 10^{-3}$ & $8.06 \times 10^{-3}$ & $1.03 \times 10^{-2}$ & $6.62 \times 10^{-3}$ & $1.70 \times 10^{-3}$ & $1.70 \times 10^{-3}$ \\
Kepler-30 & $3.53 \times 10^{-2}$ & $5.74 \times 10^{-2}$ & 0.20 & $7.58 \times 10^{-2}$ & $3.81 \times 10^{-2}$ & $3.68 \times 10^{-2}$ & $4.29 \times 10^{-2}$ & 0.14 & $4.18 \times 10^{-2}$ & $4.10 \times 10^{-2}$ \\
\hline
\end{tabular}
\label{table_rvalues}
\end{center}
\end{tiny}
\end{table*}

\begin{table*}
\begin{center}
\caption{Results of MCMC analysis of the intervals in Table~\ref{table2cc} for the stars of our sample that show evidence of DR.}
\begin{tabular}{ccccccc}
\hline
Star name & $P_{1}$ & $\sigma_{P_{1}}$ & $P_{2}$ & $\sigma_{P_{2}}$ & $\Delta P /P$ & $\sigma_{\Delta P}/P$ \\
 & (d) & (d) & (d) & (d) \\
 \hline 
 $\epsilon$ Eridani & 11.3592 & $3.181 \times 10^{-2}$ & 11.5593 & $1.853 \times 10^{-2}$ & 0.0175 & $3.21 \times 10^{-3}$ \\
 CoRoT-6 & 7.9655 & $3.799 \times 10^{-2}$ & 6.3438 & $2.033 \times 10^{-2}$ & 0.2266 & $6.02 \times 10^{-3}$ \\
 KIC~7765135 & 2.6227 & $1.492 \times 10^{-3}$ & 2.3664 & $1.142 \times 10^{-3}$ & 0.1027 & $7.53 \times 10^{-4}$\\
 KIC~7985370 & 2.9262 & $3.912 \times 10^{-3}$ & 2.7809 & $2.468 \times 10^{-3}$ & 0.0509 & $1.62 \times 10^{-3}$\\
 KIC~8429280 & 1.20593 & $2.981 \times 10^{-5}$ & 1.17245 & $2.689 \times 10^{-5}$ & 0.0281 & $3.38 \times 10^{-5}$ \\ 
 Kepler-30 & 15.9243 & $1.708 \times 10^{-2}$ & 16.7800 & $1.945 \times 10^{-2}$ & 0.0523 & $1.58 \times 10^{-3}$\\
 \hline
\end{tabular}
\label{table3}
\end{center}
\end{table*}

\section{Conclusions}

We have introduced a method of searching for differential rotation (DR) signals in high-precision  photometric time series of the late-type stars acquired by space-borne telescopes (see Sect.~\ref{overview}). The even sampling of the time series  allowed us to apply an autocorrelation to measure the timescale of the intrinsic evolution of the spot pattern that limits the possibility of measuring DR. 
We selected a sample of eight stars for which previous determinations of  DR based on photometry were available along with some information on the inclination of the stellar rotation axis. We added the Sun as a star to this sample to study the rotation of solar analogues. 
We found that a significant DR can be detected  when  the relative height of the second maximum in the autocorrelation function is at least $0.6-0.7$.  In this case, we subdivided the time series into intervals the duration of which was shorter than the typical timescale of starspot evolution, although   long enough to reveal the drift of the spot longitudes produced by DR. The best interval was  then selected  by comparing the best fits obtained with a simple two-spot model with and without DR. For that interval, a fully Bayesian analysis was undertaken by means of an MCMC approach to assess the significance of DR and determine its most probable value and uncertainty. 

We found that the amplitude of the DR derived with our method is generally consistent with previous results  when the different assumptions of the other approaches are considered. The advantage of our approach is the simplicity of the spot model that allows us to run MCMCs with tens or hundreds of million steps to study the a posteriori distribution of the parameters. This is particularly useful in spot modelling given the strong correlations among different parameters. We took advantage of the available information on the inclination of the stellar rotation axis to fix the a priori distribution of the inclination that is strongly correlated with the colatitudes and the unprojected areas of starspots in our model. Other correlations may become important when the signal-to-noise ratio of the photometry is low, spots are rapidly evolving ($\tau_{\rm s} \la (2.5-3.0) P $), or the DR relative amplitude is small ($\Delta P / P \la 0.01$). They severely hamper the proper mixing and convergence of the MCMC procedure, although their effect can be controlled  as described in Sect.~\ref{param_corr}, provided that a clear signal of DR is present in the light curve. 

In the case of stars more active than the Sun, we find that the measured amplitude of DR depends on the specific time interval considered. In general, our approach only provides a lower limit to the amplitude of surface DR. We did not attempt to extract the amplitude and sign of the pole-equator shear as in other works \citep[e.g., ][]{Froehlichetal12} because spot colatitude and inclination of the stellar spin axis are highly correlated in our simple spot model. On the other hand, an independent estimate of the inclination is generally not available, except in the case of close eclipsing binaries or transiting star-planet systems.  

\begin{acknowledgements}
The authors are grateful to an anonymous referee for a careful reading of the manuscript and valuable comments that helped to improve their work. 
AFL is grateful to Drs. A.~S.~Bonomo, H.-E.~Fr\"ohlich, D.~Gandolfi, and Prof.~S.~Ingrassia for useful discussions. 
MLC acknowledges the CAPES Brazilian agency for a PDSE fellowship (BEX-9103/12-0) and I.~C.~Leao for providing  software for light curve treatment. Research activities of the Stellar Board of the Federal University of Rio Grande do Norte are supported by CNPq and FAPERN Brazilian agencies. JRM and MLC also acknowledges the INCT INEspa\c{c}o.
This investigation made use of data from {\it MOST} (a Canadian Space Agency mission, jointly operated by Microsatellite Systems Canada Inc. (MSCI; formerly Dynacon Inc.), the University of Toronto Institute for Aerospace Studies, and the University of British Columbia, with the assistance of the University of Vienna), {\it CoRoT} (a space mission developed and  operated by the CNES, with participation of the Science Programmes of ESA, ESA's RSSD, Austria, Belgium, Brazil, Germany, and Spain), and {\it Kepler} (a NASA space mission), the public availability of which is gratefully acknowledged. 
\end{acknowledgements}

\appendix

\section{Correlations between parameters adopted in MCMC spot modelling}
\label{app_corr}
To justify the correlations among parameters adopted in some of our MCMC modelling, we consider that they are most evident when a given spot $j$ produces its maximum photometric effect, that is when  $\mu$ as given by Eq.~(\ref{mu_def}) is at its maximum $\mu_{j}^{\rm (m)}$. This occurs at an epoch $t_{j}^{\rm (m)}$ when the phase $\lambda_{j} + \Omega_{j} (t_{j}^{\rm (m)} - t_{0}) = 0$, that is $t_{j}^{\rm (m)} - t_{0} = -\lambda_{j}/\Omega_{j}$. At that epoch we have:  
\begin{equation}
\mu_{j}^{\rm (m)} = \cos( i- \theta_{j}).
\label{app1} 
\end{equation}
From Eqs.~(\ref{app1}) and~(\ref{spot_effect}), we see that the maximum photometric effect of a given spot is not changed when $\delta \theta_{j} = \delta i$, thus establishing a strong correlation between the variations of its colatitude and of the inclination of the stellar rotation axis. 

Another correlation comes from  the phase of maximum spot visibility that after variation yields: 
\begin{equation}
\delta \lambda_{j} + \delta \Omega_{j} (t_{j}^{\rm (m)} - t_{0}) = 0, 
\end{equation}
where $t_{j}^{\rm (m)} - t_{0}$ is not varied because the epoch of maximum visibility is constrained by  a relative minimum in the flux along the observed time series. Since $t_{j}^{\rm (m)} -t_{0} = -\lambda_{j}/ \Omega_{j}$ and $\delta \Omega_{j} / \Omega_{j} = -\delta P_{j} / P_{j}$, we obtain the correlation:
\begin{equation}
\delta \lambda_{j} / \lambda_{j} = - \delta P_{j} / P_{j}
\end{equation}
between the longitude $\lambda_{j}$ and the rotation period $P_{j}$ of a given spot. 

\end{document}